\documentclass[
  journal=largetwo,
  manuscript=article-type,
  year=2024,
  volume=37,
]{cup-journal}

\usepackage{amsmath}
\usepackage[nopatch]{microtype}
\usepackage{booktabs}

\usepackage{graphicx, longtable, pdflscape, supertabular} 
\usepackage{hyperref}
\usepackage{caption,subcaption,wrapfig}
\graphicspath{{figures/}}
\usepackage{anyfontsize} 
\usepackage{graphicx}	
\usepackage{amsmath}	
\usepackage{subcaption}
\usepackage{multirow}
\usepackage{comment}
\newcommand\sun{\hbox{$\odot$}}

\title[S Ant \& $\epsilon$ CrA]{Comparative study of the W UMa type binaries S Ant and  $\epsilon$ CrA}

\author{Volkan Bak{\i}\c{s}}
\affiliation{Department of Space Sciences and Technologies, Faculty of Sciences, Akdeniz University, 07058 Antalya, Turkey}
\email[Volkan Bak{\i}\c{s}]{volkanbakis@akdeniz.edu.tr}

\author{Edwin Budding}
\affiliation{Carter Observatory, {40 Salamanca Road, Kelburn, Wellington 6012, New Zealand}}
\alsoaffiliation{School of Chemical \& Physical Sciences, Victoria University of Wellington, PO Box 600, Wellington 6140, New Zealand}

\author{Ahmet Erdem}
\affiliation{Astrophysics Research Center \& Ulupınar Observatory, \c{C}anakkale Onsekiz Mart University, TR-17100, Çanakkale, Turkey}

\author{Tom Love}
\affiliation{Variable Stars South, RASNZ, PO Box 3181, Wellington 6011, New Zealand}

\author{Mark G. Blackford}
\affiliation{Variable Stars South, Congarinni Observatory, Congarinni, NSW, 2447, Australia}

\author{Wu Zihao}
\affiliation{Dept.\ Statistics \& Data Science, National University of Singapore, 6 Science Drive 2, Singapore 117546}

\author{Adam Tang}
\affiliation{University of Chicago Laboratory Schools, 1362 E 59th St, Chicago, IL 60637, USA}

\author{Michael D. Rhodes}
\affiliation{{Brigham Young University, Provo, Utah 84602,} USA}

\author{Timothy S. Banks}
\affiliation{Nielsen, 675 6th Ave, New York, NY 10011, USA}
\alsoaffiliation{Dept.\ of Physical Science \& Engineering, Harper College, 1200 W Algonquin Rd, Palatine, Illinois 60067, USA}


\begin{document}

\begin{abstract} 
Contact binaries challenge contemporary stellar astrophysics with respect to their incidence, structure and evolution.  We explore these issues through a detailed study of two bright examples: S Ant and $\epsilon$  CrA, that permit high-resolution spectroscopy at a relatively good S/N ratio. The availability of high-quality photometry, including data from the TESS satellite as well as Gaia parallaxes, allows us to apply the Russell paradigm to produce reliable up-to-date information on the physical properties of these binaries.  As a result, models of their interactive evolution, such as the thermal relaxation oscillator scenario, can be examined. Mass transfer between the components is clearly evidenced, but the variability of the O'Connell effect over relatively short time scales points to irregularities in the mass transfer or accretion processes. Our findings indicate that S Ant may evolve into an R CMa type Algol, while the low mass ratio of $\epsilon$ CrA suggests a likely merger of its components in the not-too-distant future.   
  
\end{abstract}



\section{Introduction} \label{sec:intro}

The close eclipsing binary stars S Ant and $\epsilon$ CrA are examples of the relatively plentiful `contact' (sometimes {\em overcontact}) systems associated with the smoothly varying EW, or W UMa, class of light curve (LC).  Within this grouping,  S Ant and $\epsilon$ CrA are of  \citeauthor{Binnendijk_1966}'s (\citeyear{Binnendijk_1966})  `A' type, or class,  characterized by the primary minimum resulting from the eclipse of the bigger, more massive, and higher mean surface temperature component.  This is distinguished from the more frequently found `W' type, where the primary minimum is caused by the eclipse of the smaller, hotter, star \autocite{Binnendijk_1966}. 

$\epsilon$ Coronae Australis is the brightest known example of the W UMa stars.  \cite{Eker_2009} gave the mass ratio ($q$) of $\epsilon$ CrA as 0.128, which is very similar to that found by \cite{Goecking_1993}.  The mass ratio was recently confirmed in the detailed spectroscopic analysis of \cite{Rucinski_2020}. This low secondary mass at once points to peculiarity given that the component luminosities are not greatly different from each other.  The mass ratio for the comparable system S Ant is also fairly low at $q = 0.33 \pm 0.02$ \citep{Duerbeck_2007}. S Ant is similarly a bright A class system, also well to the south in declination. The availability of good signal-to-noise ratios in a high spectral resolution context, along with indications that both binaries are engaged in a mass transfer process,  thus commends attention to these relatively neglected objects.

It has been generally taken that the outer layers of binaries like S Ant and $\epsilon$ CrA are in physical contact, with a common envelope surrounding two mass concentrations. Such a model explains the general shape of the LC, which has to involve relatively close proximity of the two mass centres given the strong tidal distortions evident from the markedly changing received light fluxes with the orbital phase. The common envelope may then account for the comparability of the two luminosities on the basis of energy transferred between the underlying stars \citep{Mochnacki_1981, Hilditch_1988}.  Such a picture has been discussed for decades, with the availability of decisive evidence generally insufficient to unequivocally settle the modelling (see \citeauthor{Yakut_2005}, \citeyear{Yakut_2005}, and the many references therein). The direct association of physical characteristics with LC type is ambiguous, particularly with the smooth, quasi-sinusoidal EW LCs, where physically different configurations may show similar forms of photometric variation. Analyses of such LCs indicate that the components have dimensions that would put them close to theoretical surfaces of limiting stability, but some such binaries might just be detached pairs whose components happen to be relatively close to this limit; for example, V831 Cen \citep{Budding_2010}.

\begin{figure}[t]
    \centering
    \includegraphics[width=1.0\columnwidth]{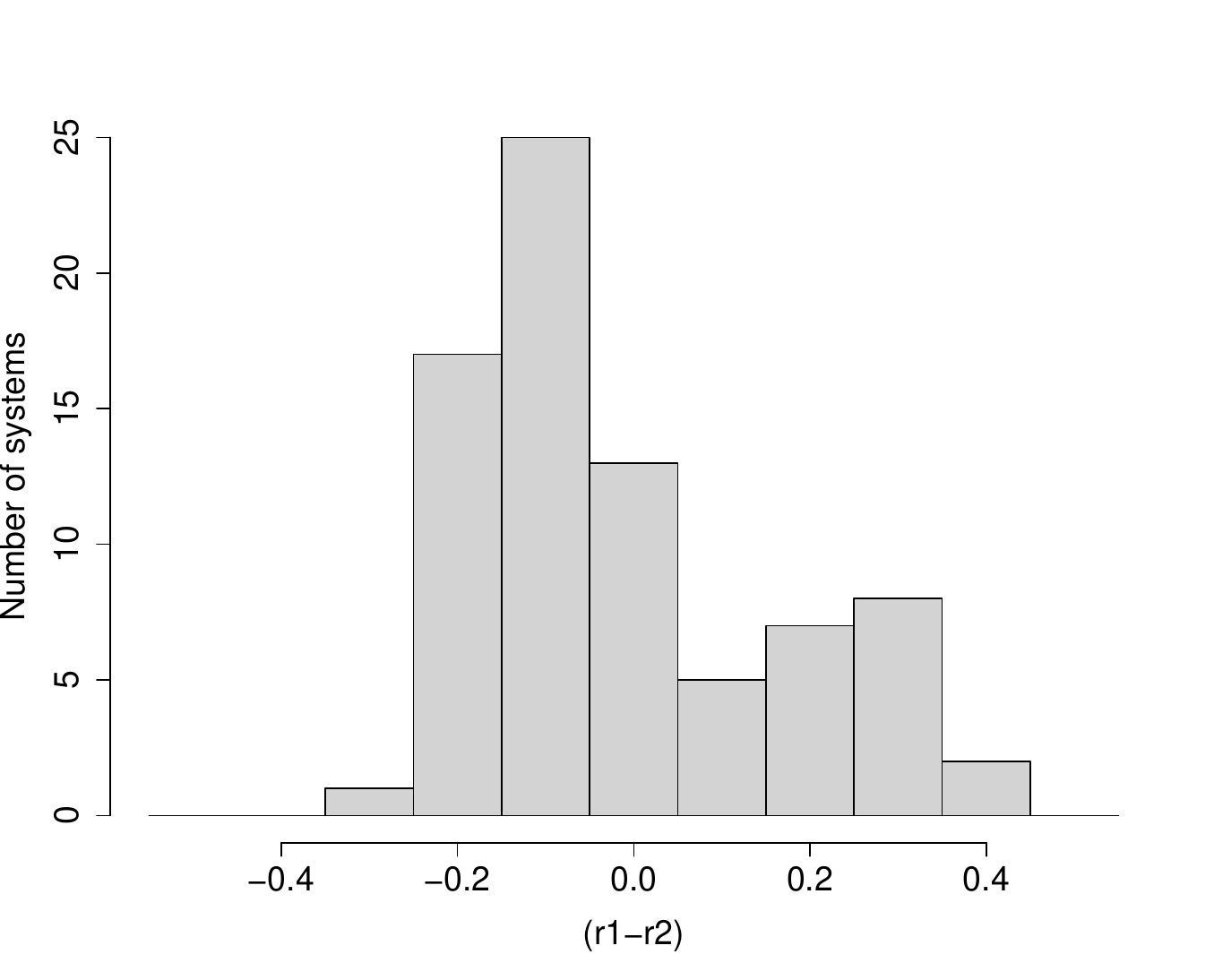}
    \vspace{-0.5cm}
    \caption{\label{fig:relative_radii} Distribution of differences in the relative radii of 78 well-studied EW-type binaries from the catalogue of \protect\cite{Maceroni_1996}. $r_1$ is the radius of the hotter star, as a fraction of the separation of the two mass centres. The A-type systems are to the right and the  W types are to the left of the central axis.}
\end{figure}

In Fig~\ref{fig:relative_radii} we plot the frequencies of the difference in relative radii $r_1 - r_2$ (i.e.\  stellar mean radius divided by the separation of the mass centres) from \cite{Maceroni_1996}'s catalogue of well-studied contact binaries. Here we set $r_1$ to correspond to the hotter component: the one eclipsed at primary minimum, so A-type systems are on the right of the origin and W-types on the left. This difference is also a measure of the difference in the mass ratio $q$ (usually given in the sense mass$_{\rm smaller}$/mass$_{\rm greater}$)  from unity, as, for contact binaries, the relative radii are fixed by $q$.\footnote{In fact, $r_2/r_1 \approx q^{0.46}$ \citep{Kuiper_1941}, while $r_1 + r_2 \approx 0.75$ \citep{Kopal_1959}.} 

It should be noted that here we are using the `photometric' identification of $r_1$ as the relative radius of the star of greater surface brightness:  the star eclipsed at the primary minimum, as in the catalogue of \cite{Maceroni_1996}. In spectroscopic contexts the 'primary'  normally refers to the more massive star, These spectroscopic and photometric primary stars are not necessarily the same object.  For contact binaries there is the added constraint that the larger star is the more massive one, so if it is also the hotter one we must be dealing with an A-type system, having $r_1 > r_2$.  If the hotter star is smaller, $r_1 < r_2$ corresponding to the W type.   \cite{Binnendijk_1966} followed the spectroscopic convention with $r_1$ applying to the more massive star, and so $q  < 1$. \cite{Binnendijk_1966}  also had (in his notation) $L_1 >  L_2$,  as an empirical finding. That would be certainly true for the A type systems.  For  W type systems, since $T_1$ and $T_2$ are usually quite similar, if $r_1$ is significantly bigger then $r_2$ then  $L_1 >  L_2$   would be likely also for the W-types, but this is not rigorously established in general.

The distribution in Fig~\ref{fig:relative_radii} shows the existence of the two distinct clumps of W and A types, with a general prevalence of $q$-values not too far from unity.\footnote{ The non-parametric Wilcoxon test supports that the two groupings shown in Fig~\ref{fig:relative_radii} are statistically different at greater than 99\% confidence. Before conducting a Welch two-sample t-test (which assumes that the tested populations have normal distributions), we conducted Shapiro-Wilk normality tests of the A and W subpopulations. Both these tests were significant at greater than 99\% confidence, as was the final t-test. We therefore conclude that the two distributions are indeed distinctly separated on the chosen variable.} It suggests some basic differences in the mechanisms leading up to the W and A types rather than a single continuum of formation see Fig ~\ref{fig:relative_radii}. 

Extreme mass ratios disappear from the population relatively quickly, and none are found below a certain limiting mass ratio.  This point, although possibly influenced by selection effects, was noted by \cite{vant_Veer_1994}, and discussed by \cite{Rasio_1995} and \cite{Li_2006}, who showed that contact binary models would be subject to dynamical instability if the mass ratio (lesser/greater) was less than about 0.09. This is slightly greater than the value 0.075 cited by \cite{Rucinski_1972} for AW UMa. The predicted result of the mass ratio falling below the limit is a merger of the components into one rapidly rotating object, V1309 Sco being an example \citep{Tylenda_2011}. If we use the fact that $r_1 + r_2 \approx 0.75$ \citep{Kopal_1959}, then the drop-off in the distribution at $|r_1 - r_2| \approx 0.3$ supports a swift decline in the number of contact binaries when $q$ drops below about 0.15. Keeping in mind observational uncertainties, the absence of systems for which $|r_1 - r_2| > 0.5$ confirms the absence of systems in this sample for which $q$ is less than at least 0.05, although this was recently challenged by \cite{Li_2023}.

In another development, \cite{Stepien_2009} considered a mass transfer, as generally recognized in the Algol-type binaries. Such a model posits the present secondary in W UMa systems to be at a more advanced evolutionary stage than the primary.  The initially Main-Sequence-like primary expands beyond its critical surface so as to generate a mass-transferring stream. Coriolis forces deflect the stream from the line of centres of the two stars as well as preventing matter from the primary attaining a high latitude on the secondary star.  The movement of mass is confined to a central zone about the equator.  In this scheme, the stream, with its associated thermal energy, circulates around the secondary and may then return to the primary. 
The model predicts that spectroscopic observations should be able to detect this flow.  Observations consistent with such motions were reported by Rucinski in two cases: AW UMa \citep{Rucinski_2015}, and $\epsilon$ CrA \citep{Rucinski_2020}.  
Substantiation of this picture forms a clear motivation for the present study.

Short literature reviews of the two binary systems S Ant and $\epsilon$ CrA follow.   Given the importance of the mass ratio to the parametrization of contact binaries, the introduction is followed by the spectrometric section that determines the mass ratio from the radial velocity (RV) data.  We present and analyse new photometry in Section~\ref{sec:photometry}. Indications of trends of period variation for both binaries follow in Section~\ref{sec:period}. Absolute parameters are given in Section~\ref{sec:absolute_parameters} and we discuss relevant issues in the final section.

\subsection {S Antliae}

S Ant (= HD 62810; HIP 46810, HR 3798) is a mag V $\approx$ 6.5, B -- V $\approx$ 0.32; $\lambda$ $\approx$ 258.5, $\beta$ $\approx$ +16.6 (SIMBAD, \citeauthor{Wenger_2000}, \citeyear{Wenger_2000}) F3V dwarf eclipsing binary with a short period ($P \approx 0.6483$ d).  The period appears to be slowly increasing \citep{Kreiner_2001}, suggesting a long-term widening out of the two stars, to preserve angular momentum,  with mass transfer from the secondary in a `semi-detached' state. In saying that, it is noteworthy that various physical factors affect the stellar interaction processes, including magnetic braking, short-term flow instabilities or dissipative effects.  Without additional evidence, the interpretation of apparent period variation over a relatively short time interval remains speculative.


S Ant has been the subject of various studies since its discovery   \citep{Paul}, though the star has been considered a `difficult' object (\citealt{Joy_1926, Popper_1956}) and received relatively limited spectroscopic attention hitherto.   \cite{Joy_1926} observed  S Ant between 1917 and 1921 from Mount Wilson and confirmed its identification as an eclipsing binary.  A further 26 spectra were obtained from Mount Wilson in 1954, but researchers struggled to identify features other than the hydrogen Balmer lines of the primary, due to the high degree of rotational broadening and blending  \citep{Popper_1956}. 

\cite{Duerbeck_2007} produced RV amplitudes for S Ant of $K_1 =  77.8$,  $K_2 = 234.1$, and $V_{\gamma}$ = 26.2 km s$^{-1}$, with error estimates of about 3 km s$^{-1}$, on the basis of a small number of reliable data-points (6 for the primary and 4 for the secondary) that were gathered in 1996 with the ESO 1.52m telescope at La Silla. This is significantly different from the solution of \cite{Popper_1956}, who claimed $K_1 = 92.3$ and $V_{\gamma} = -1.2$ km/s, with uncertainty  estimates of about 1 km s$^{-1}$.  \cite{Duerbeck_2007} referred to the previous photometric analysis of \cite{Russo_1982}, who found inconsistencies with the results of their LC modelling using the {\sc wd} code of \cite{Wilson_1971}.  Influenced by \citeauthor{Popper_1956}'s spectroscopic results, \citeauthor{Russo_1982} reported a photometric mass-ratio of $0.59 \pm 0.02$, though this is significantly at odds with the mass ratio given by \citeauthor{Duerbeck_2007}  ($q = 0.33 \pm 0.02$). The latter mentioned a ``notorious unreliability'' of contact binary mass ratios estimated from photometry alone. \citeauthor{Duerbeck_2007} preferred the later spectral type of F3V assigned by \cite{Houk_1982}, rather than the A8 classification of \cite{Joy_1926}, recalled by \cite{Abt_2005}.   The use of their broadening function technique to determine more reliable RV values is noteworthy ({\bf see} the papers of \citealt{Rucinski_2015, Rucinski_2020} dealing with AW UMa and $\epsilon$ CrA).

One aim of our present work is to report spectroscopic data of suitable quality that will improve the parametrization of both S Ant and $\epsilon$ CrA. To this end, these data are combined with high-precision photometry from the TESS satellite \citep{Ricker_etal_2015}, as well as supporting ground-based BVI photometry. 


\subsection {
\protect\texorpdfstring{$\epsilon$}{} Corona Australis}

\label{subsec:eps_cra_intro}

\noindent
$\epsilon$ CrA  (= HD 175813; HR 7152; mag V $\approx$ 4.8, B-V $\approx$ 0.36, P $\approx$ 0.591 d) is a bright F4V type star, at a distance of about 31 pc. The binary is reported to have a relatively high systemic velocity of close to 58 km s$^{-1}$ (SIMBAD,\citeauthor{Wenger_2000},  \citeyear{Wenger_2000}).  It has received rather less published attention than S Ant, although sharing similar A-class contact binary system properties.  \cite{Cousins_1950} observed its eclipsing binary nature and later produced a report  \citep{Cousins_1964}. The LCs of \cite{Knipe_1967} and \cite{Tapia_1969} were modelled by \cite{Twigg_1979} and \cite{Wilson_2011}, both finding an orbital inclination of $\sim 73$ degrees.  These papers adopted \cite{Lucy0}'s prescription for convective-model gravity darkening coefficients, playing down the $\sim 6500$ K surface temperature indications. \cite{Twigg_1979} estimated the mass ratio as $0.114 \pm 0.003$ and \cite{Wilson_2011} as $0.1244 \pm 0.0014$. \cite{Tapia_1975}  had earlier undertaken a radial velocity analysis on the basis of over 70 spectrograms obtained in good weather conditions at Cerro Tololo in 1972. \cite{Goecking_1993}, who gave an essentially similar mass ratio at $0.128 \pm 0.014$, used  55 Coude spectrograms taken with the 1.5-m telescope at La Silla Observatory, Chile. 
 
\cite{Rucinski_2020} presented a detailed analysis of the system, in which he argued that the \cite{Stepien_1995} model offers an improved explanation of the spectroscopic results.  Rucinski's high-quality data on $\epsilon$ CrA were obtained using the CHIRON spectrograph on the CTIO/SMARTS 1.5-m telescope over 8 nights in July 2018, during somewhat variable weather conditions. Some 361 exposures were made with the high resolution of $\sim$80,000. The spectral window was arranged to be from 506.05 to 529.20 nm, so centred on the Mg I b triplet.   Rucinski commented on apparent discrepancies with conventional contact binary models, noting a rapid, non-synchronized rotation of the primary component, apparently unaffected by the secondary. That component appears surrounded by a complex gas flow that contravenes the equilibrium requirements of Roche surfaces. 


\section{Spectroscopy}
\label{sec:spectroscopy}
\subsection{S Ant}
\label{sec:S_Ant_spectrocopy}

\begin{figure}[t]
    \centering
    \includegraphics[width=0.9\textwidth]{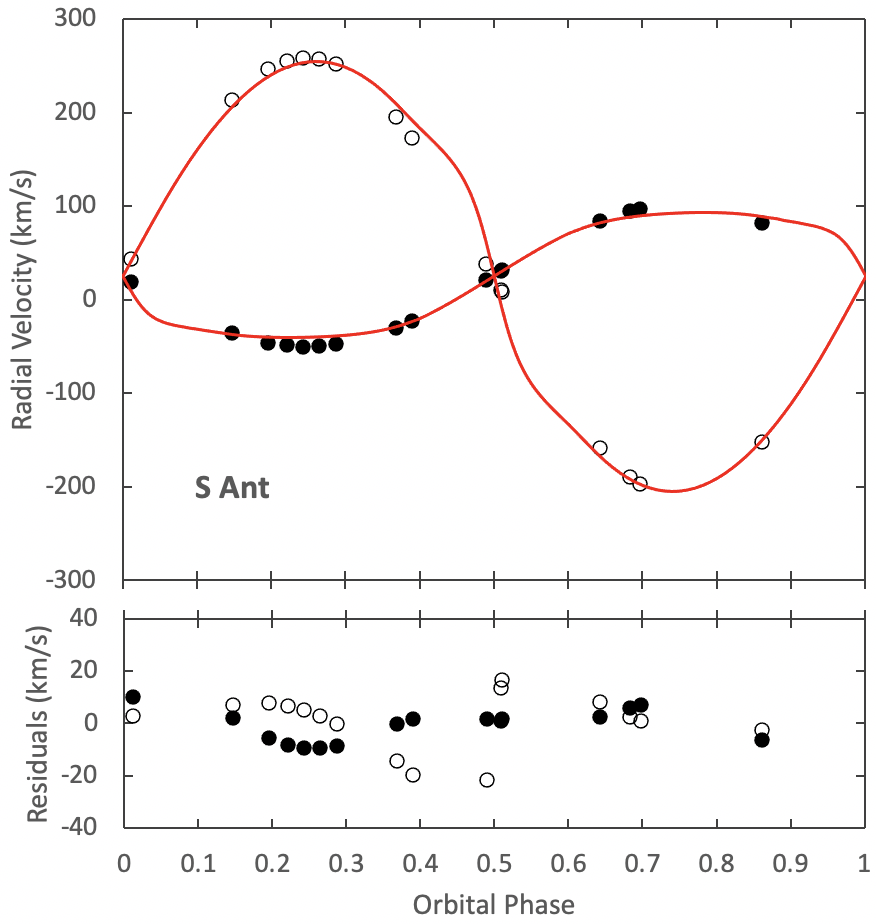}
\caption{Radial velocity curves of S Ant from {\sc Korel} applied to UCMJO data. Residuals to the RV model were plotted in the bottom figure. RVs of the primary and the secondary components are marked as filled and hollow symbols, respectively. Orbital phases were calculated using the quadratic ephemeris given in Table~\ref{table:S_Ant_o_c}.}
    \label{fig:SAntMJUO_RV}
\end{figure}

Spectroscopic data on S Ant were gathered over 5 nights during the period  January 21-26  2021 with the High Efficiency and Resolution Canterbury University Large \'{E}chelle Spectrograph (HERCULES) of the Department of Physics and Astronomy, University of Canterbury  \citeauthor{Hearnshaw_2002}, \citeyear{Hearnshaw_2002}). This was attached to the 1m McLellan telescope at the Mt John Observatory (UCMJO).  Images were collected with a 4k$\times$4k Spectral Instruments (SITe) camera \citep{Skuljan_2004}. The 100 $\mu$m fibre, which is suited to typical seeing conditions at Mt John, enables a theoretical resolution of $\sim$40,000. Exposure times were usually $\sim$900 seconds for this 6th mag star. Raw observations were reduced with the latest version of the software {\sc hrsp} \citep{Skuljan_2021}, that produces wavelength calibrated and normalized output conveniently in `FITS'\footnote{See https://fits.gsfc.nasa.gov/standard40/fits\_standard40aa-le.pdf for further details on this file format.} formatted files.  Each recording covers the wavelength range 4520-6810 {\AA}. The log of observations is given in Table \ref{tab:SAnt_obs_log} and a descriptive summary of the observed spectra, giving line identifications and suitability for measurement, is presented as Table~\ref{tab:SAnt_spectrum} (which is placed into the appendix so as not to disrupt the paper flow). 
Many metallic absorption lines are present  in these spectra, though they are considerably widened by rotational effects and often blended.  While the primary star's lines are readily located it is often difficult to identify clearly those of the secondary that have low signal-to-noise ratios and may be relatively more affected by material motions in the source medium \citep{Rucinski_2020}.   

\begin{figure}
    \centering 
    \begin{subfigure}{0.9\textwidth}
        \includegraphics[width=\linewidth]{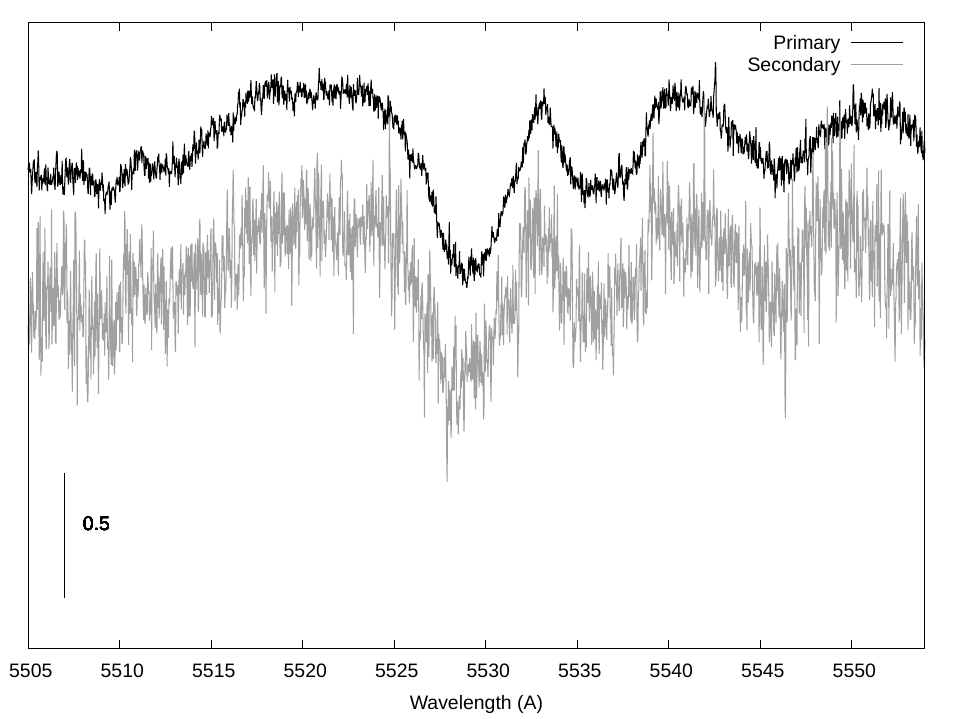}
        \label{fig:o103_reconstructed}
    \end{subfigure} \\
    \begin{subfigure}{1.\textwidth}
        \includegraphics[width=\linewidth]{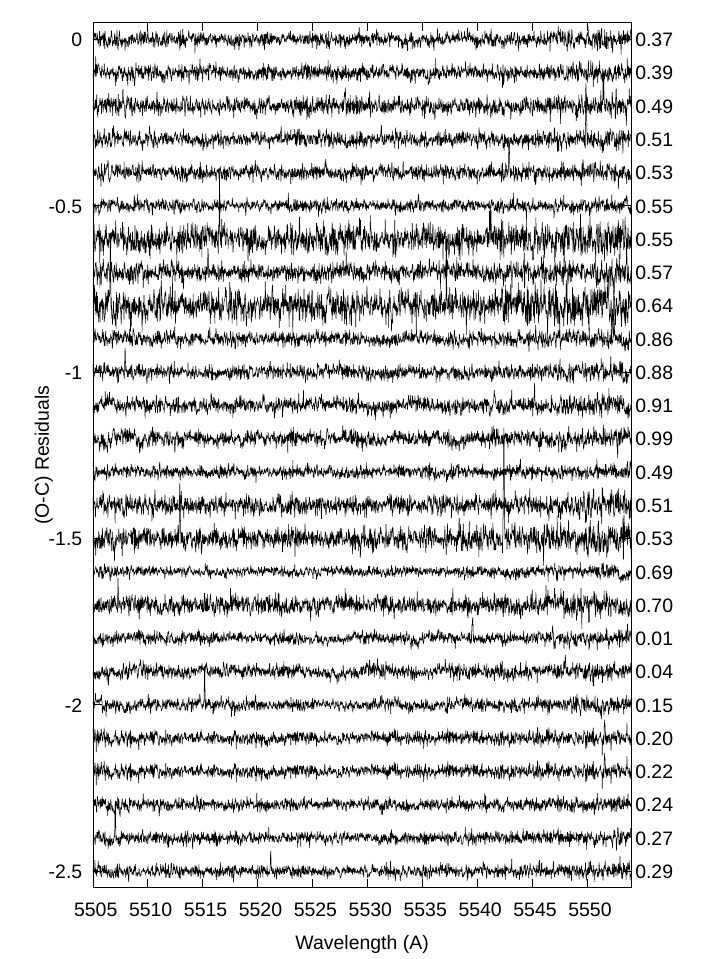}
        \label{fig:o103_OC}
    \end{subfigure}
    \caption{Disentangled spectra of the S Ant components (\textit{top}), together with  residuals (\textit{bottom}). Numbers on the right side of the residual plot refer to the orbital phase.
    \label{fig:SAnt_disentangled}}
\end{figure}

We checked the spectral type designation using our measurements of the H$_{\beta}$ equivalent width.  The value was 4.9 \AA \, -- which would place the spectral class a little cooler than F3V, according to \cite{Jaschek_1987}, but within one type-unit. 
Measurement of the B -- V colour  (section~\ref{sec:S_Ant_photometry}) agrees with this spectral type,  after a small correction for interstellar reddening. 

For two of the observations close to orbital phase 0.5 (w9236044s, w9240025s, see Table~\ref{tab:SAnt_obs_log}), the secondary component was totally eclipsed, allowing us to treat those spectral images as the primary component only. They could thus be directly modelled with synthetic spectra calculated directly with appropriate model atmosphere parameters. Fixing the surface gravitational acceleration of the primary component at log g=3.98 (cgs), and using mass, radius and composition estimates anticipated for the primary component, synthetic spectra from Kurucz model atmospheres could be constructed for a grid of effective temperature and projected rotation values to compare with the observations. Preliminary findings imply a temperature of the primary component of 7100 K and its projected equatorial surface rotational velocity of 150 km/s. The temperature of the primary component is in good agreement with its B -- V colour (section~\ref{sec:S_Ant_photometry})  and its H$_{\beta}$ equivalent width obtained above. The models are plotted on the observations in Fig.\ref{fig:SAnt_eclipse_spectrum}.

Sixteen spectra around the elongation phases of S Ant, from observations collected in clear weather, have been selected for RV analysis. The highly broadened lines did not completely separate, even at the greatest observed elongation.

As a check on procedure, the data of \cite{Duerbeck_2007} were analysed using the {\sc ELISa} \citep{Cokina_2021} package for parametrization, and a Markov Chain Monte Carlo (MCMC) technique for the curve-fitting process. {\sc ELISa} includes both LC and RV analytical components. The {\sc ELISa} search for an optimal parameter set depends on Markov Chain sampling of parameter hyperspace, while the probability estimation invokes an essentially random, `Monte Carlo', distribution of observational errors. 

With preliminary orbital parameters obtained by {\sc ELISa}, we applied the spectral disentangling technique {\sc korel} \citep{hadrava95} to the observed spectra of S Ant. Spectra during eclipses were excluded, except for one mid-eclipse spectrum, where the Rossiter-McLoughlin effect \citep{Rossiter_1924, McLaughlin_1924} vanishes. The mean relative light contributions of the components were taken from the results of our initial LC analysis (next Section). A total of 8 spectral orders (93, 97-103) were processed with {\sc korel}. The derived RVs from each order have been averaged to produce the final set. The final RVs are presented in Table~\ref{tab:S_Ant_obs_log_1} and displayed in Fig~\ref{fig:SAntMJUO_RV}. In Table~\ref{tab:mjuo_rv_fit} we present orbital parameters from the spectral disentangling procedure, where symbols have their usual meanings. Uncertainty estimates are produced as part of the output of the fitting program.
 
In Fig. \ref{fig:SAnt_disentangled}, we display reconstructed mean spectra of the strong, measurable features in the $\lambda$552-559 nm region of the component stars of S Ant, together with residuals from the {\sc korel} fits. It appears from the residual plots, that there is no significant additional component in these spectra. 
As well, profile fitting using the program {\sc prof} (see, e.g., \cite{Olah_1992}) enabled us to determine Doppler shifts of the relatively well-defined $\lambda$4824.13 Cr II line to be determined in a consistent way.
In Table~\ref{tab:mjuo_rv_fit} we also present orbital parameters from the RV data obtained with the program {\sc prof}.

\begin{table}[t]
\begin{center}
\scriptsize
 \caption{Radial velocity measurements of the components of S Ant from {\sc Korel} applied to UCMJO data. The values in the $O-C$ columns represent the deviations of the individual measurements from the fitted RV curves. Orbital phases were calculated using the quadratic ephemeris given in Table \ref{table:S_Ant_o_c}. }
    \label{tab:S_Ant_obs_log_1}
    \begin{tabular}{ccrrrr}
\hline
Time        & Phase     & RV$_{1}$  & (O-C)$_{1}$   &   RV$_{2}$    &   (O-C)$_{2}$ \\
(HJD-2450000)   &       & (km s$^{-1}$) & (km s$^{-1}$) & (km s$^{-1}$) & (km s$^{-1}$)	 \\
\hline
59236.0228	&	0.3681	&	$-29.8$	&	$-0.3$	&	195.8	     &	$-14.3$	  \\
59236.0367	&	0.3896	&	$-22.3$	&	1.5	    &	172.7	     &	$-19.6$	  \\
59236.1018	&	0.4899	&	21.5	&	1.6	    &	38.5	     &	$-21.8$	  \\
59236.1152	&	0.5106	&	31.7	&	1.6	    &	7.9	         &	16.5	  \\
59238.1458	&	0.6425	&	84.5	&	2.4	    &	$-158.9$	 &	8.4	      \\
59238.9358	&	0.8611	&	82.4	&	$-6.4$	&	$-152.3$	 &	$-2.5$	  \\
59240.0042	&	0.5090	&	30.4	&	1.1	    &	10.2	     &	13.6	  \\
59240.1173	&	0.6833	&	94.3	&	5.9	    &	$-189.5$	 &	2.5	      \\
59240.1264	&	0.6975	&	96.7	&	6.9	    &	$-196.8$	 &	0.8	      \\
59240.9783	&	0.0113	&	19.2	&	10.0	&	43.8	     &	2.8	      \\
59241.0664	&	0.1473	&	$-35.5$	&	2.2	    &	213.5	     &	7.2	      \\
59241.0976	&	0.1954	&	$-46.0$	&	$-5.4$	&	246.3	     &	7.8	      \\
59241.1142	&	0.2210	&	$-49.0$	&	$-8.1$	&	255.6	     &	6.6	      \\
59241.1285	&	0.2431	&	$-50.1$	&	$-9.3$	&	258.9	     &	5.0	      \\
59241.1427	&	0.2650	&	$-49.7$	&	$-9.4$	&	257.8	     &	2.7	      \\
59241.1575	&	0.2877	&	$-47.8$	&	$-8.5$	&	251.9	     &	$-0.3$	   \\
\hline
    \end{tabular}
\end{center}
\end{table}

\begin{table}
\begin{center}
\scriptsize
\caption{S Ant: Fitting parameters for RV data from UCMJO. $V_{\gamma}$ value of {\sc Korel} solution has been adopted from individual RV measurements.}
\label{tab:mjuo_rv_fit}
\begin{tabular}{lll}
\hline 
Parameter                     & {\sc prof} Value          & {\sc Korel} Value    \\
\hline                              
$P$ (days)                    & {0.64834368 (fixed)}      &  {0.64834935 (fixed)}   \\
$T_0$ (HJD-2,450,000)         & 35139.9553 (0.0004)       & 58518.7088 (0.0001) \\
$K_1$  (km s$^{-1}$)          & 80 (10)                   & 75 (1)   \\
$K_2$  (km s$^{-1}$)          & 240 (5)                   & 234 (1)    \\
$V_{\gamma}$  (km s$^{-1}$)   & 18.7 (5.1)                & 25.3 (0.1)   \\
$\Delta \phi_0$ (deg)         & 5.1 (1.2)                 & -- \\
$a \sin i$                    & 0.0184 (0.001) AU         & 3.96 (0.01) R$_{\odot}$ \\
$(M_1+ M_2)\sin^3i$ \:(M$_{\odot}$)        
                              & 2.0 (0.02)                & 1.99 (0.01)   \\
\hline  
\end{tabular}
\end{center}
\end{table}


The parameters listed in Table~\ref{tab:mjuo_rv_fit}, apart from $K_1$ that derives from direct measurements, depend on constraints imposed by the physical circumstances. Thus, the mass function $f$ (neglecting the eccentricity term and with the constant $C = 1.03615\times 10^{-7}$) yields:
\begin{equation}
    f = CP K_1^3 = \frac{M_1 q^3 \sin^3 i}{(1+q)^2} .
\end{equation}
This can be re-written as: 
\begin{equation}
    K_2 = \sin i \left( \frac{M_1}{CP} \right)^{1/3}
 (  1 + q)^{-2/3} \,  ,
\end{equation}
where the right side becomes slowly varying for low $q$, with $M_1$  constrained by the spectral type.  The result is that $K_2 \approx 235 \pm 5$ km s$^{-1}$, for plausible values of $q$ in the range 0.25-0.35 and primary RV amplitudes in the range 65-85 km s$^{-1}$. Values of $K_2$ inferred from direct measurements are compatible with this but have a much greater probability of error.


\begin{figure}
    \centering
    \includegraphics[width=0.9\textwidth]{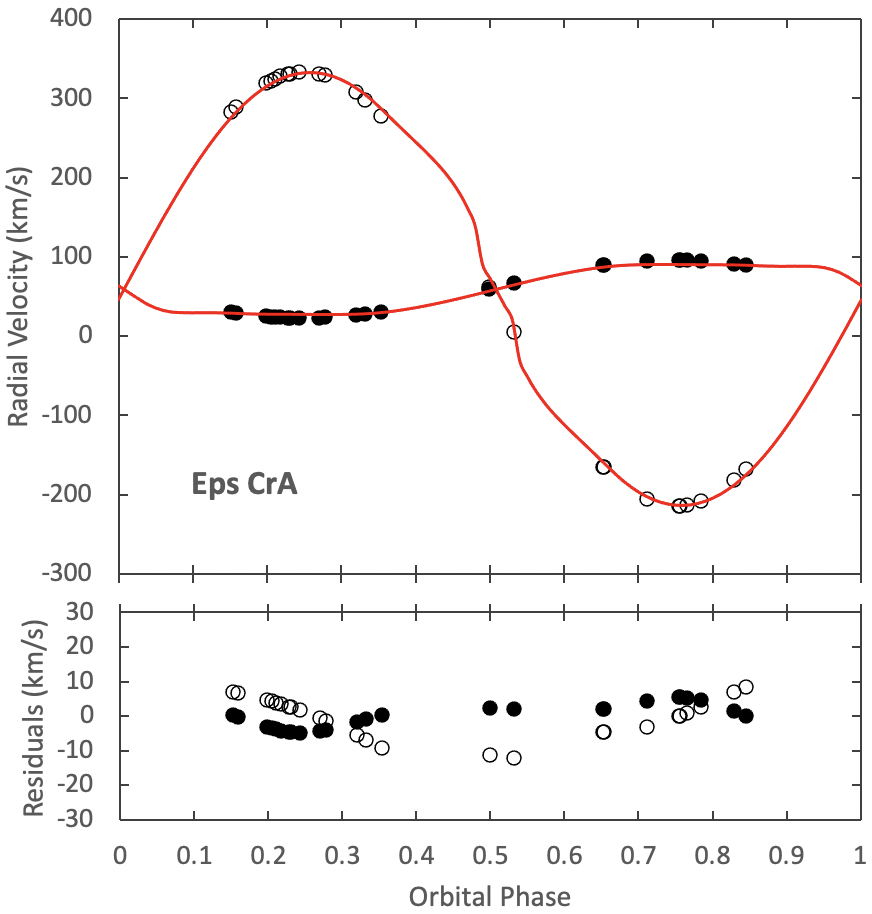}
\caption{Radial velocity curves of $\epsilon$ CrA from {\sc Korel} applied to UCMJO data. Residuals to the RV model were plotted in the bottom figure. RVs of the primary and the secondary components are marked as filled and hollow symbols, respectively. Orbital phases were calculated using the linear ephemeris given in Table \ref{tab:korel_pars}.}
    \label{fig:eps_cra_rvc}
\end{figure}

\begin{figure}
    \centering 
    \begin{subfigure}{0.9\textwidth}
        \includegraphics[width=\linewidth]{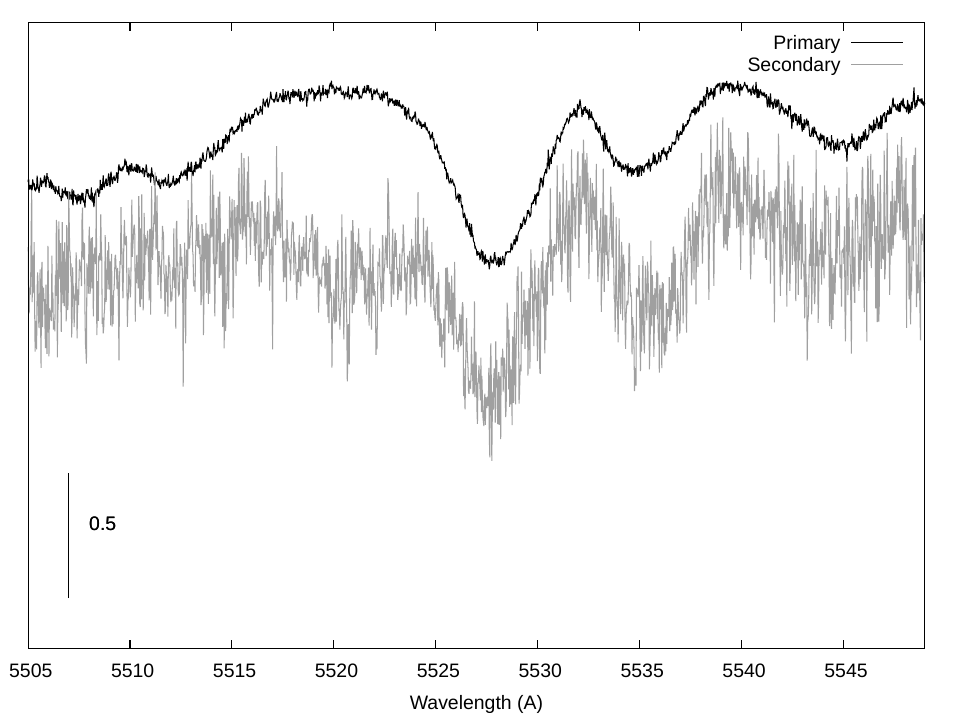}
        \label{fig:o104_reconstructed}
    \end{subfigure}
    \begin{subfigure}{1.\textwidth}
        \includegraphics[width=\linewidth]{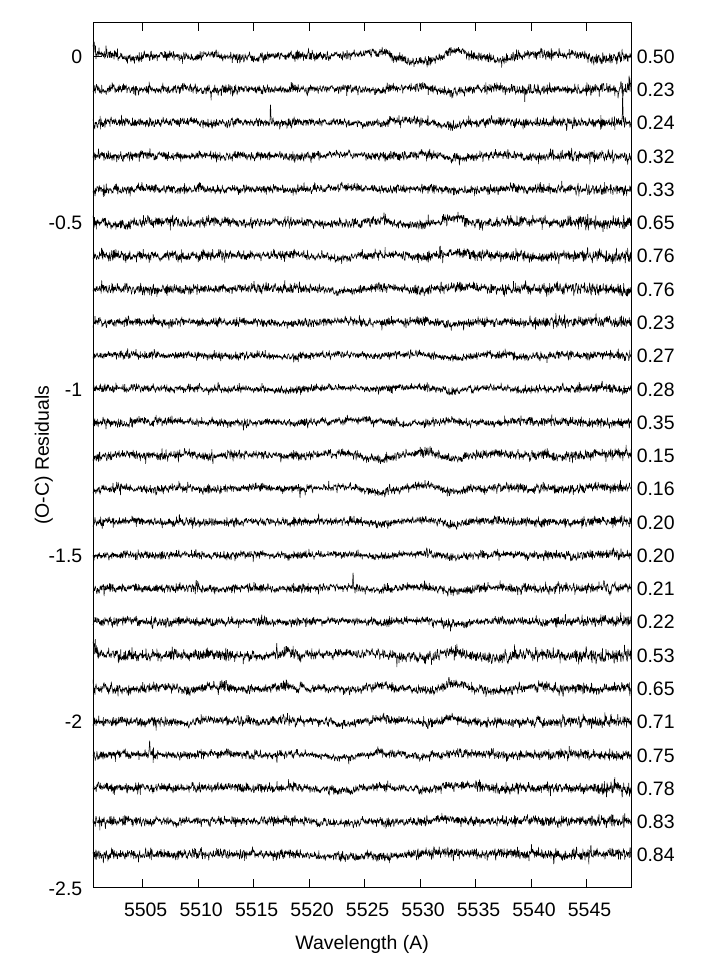}
        \label{fig:o104_OC}
    \end{subfigure}
\caption{Disentangled spectra of the $\epsilon$CrA components (\textit{top}), together with residuals (\textit{bottom}). Numbers on the right side of the O--C plot refer to the orbital phase. }
    \label{fig:epsCrA_disentangled}
\end{figure}

\subsection{\protect\texorpdfstring{$\epsilon$}{} CrA} 
\label{sec:eps_cra_spectroscopy}

The spectrographic data for $\epsilon$ CrA were obtained with essentially similar UCMJO arrangements as reported above for S Ant, except that the data for $\epsilon$ CrA were obtained in August 2006.   The log of observations is given in Table~\ref{tab:epsCrA_obs_log}. See Table~\ref{tab:epscra_spectrum} (in the appendix) for a discussion of key features in the spectra.  

We started the analysis by trial-fitting the radial velocities (RVs) of \cite{Rucinski_2020}, adopting Ruci\'{n}ski's period and applying again the  {\sc ELISa} program \citep{Cokina_2021}. We found  very similar results to those  of \cite{Rucinski_2020}, i.e.\ mass ratio $q = 0.128^{+0.005}_{-0.003}$, $a \sin{i} = 3.54 \pm 0.04$ R$_{\odot}$, and  centre of mass velocity $V_{\gamma} = 61.88^{+0.48}_{-2.59}$ km s$^{-1}$, giving  confidence to the parameterization procedures.  The high systemic velocity (\citeauthor{Fuchs_1987}, \citeyear{Fuchs_1987}) suggests a great age to the binary, in keeping with other ideas about low mass-ratio A-class contact binaries.

As in the case of S Ant, the secondary component of $\epsilon$ CrA was totally eclipsed at phase 0.5. For two of the observations close to orbital phase 0.5 (w3966049s, w3966050s, see Table~\ref{tab:epsCrA_obs_log}), we modelled with the observed spectra with synthetic spectra calculated directly with appropriate model atmosphere parameters. This time the surface gravitational acceleration of the primary component was fixed at log $g$=4.05 (cgs), and using mass, radius and composition estimates anticipated for the primary component,  synthetic spectra from Kurucz model atmospheres could be constructed for a grid of effective temperature and projected rotation values to compare with the observations. Preliminary findings are of rather low $T_{\rm eff}$ and high $V_{\rm rot}$sin$i$ values and we defer further discussion of this work in the present paper. Concerning high projected equatorial rotation velocities, however, this may align with the discussion of \cite{Rucinski_2020} on the complex velocity field in $\epsilon$ CrA associated with the effects of mass transfer. The synthetic spectra plotted over the observed spectra are shown in Fig.~\ref{fig:epsCrA_eclipse_spectrum}.

Preliminary mean relative light contributions of the components are essential for the proper decomposition of the component spectra from the composites in the source data \citep{hadrava95, hadrava97}.  The LC and RV analysis is thus, in principle, an iterative procedure, although this would seldom involve more than one parametrization cycle. 

\begin{table}[t]
\begin{center}
\scriptsize
 \caption{Radial velocity measurements of the components of $\epsilon$ CrA from {\sc Korel} applied to UCMJO data. The values in the $O-C$ columns represent the deviations of the individual measurements from the fitted RV curves. Orbital phases were calculated using the linear ephemeris given in Table \ref{tab:korel_pars}. }
    \label{tab:eps_cra_rvs}
    \begin{tabular}{cccrrr}
\hline
Time        & Phase     & RV$_{1}$  & (O-C)$_{1}$   &   RV$_{2}$    &   (O-C)$_{2}$ \\
(HJD-2450000)   &       & (km s$^{-1}$) & (km s$^{-1}$) & (km s$^{-1}$) & (km s$^{-1}$)	 \\
\hline
53966.9845	&	0.4993	&	59.5	&	2.3	    &	62.6	    &	$-11.3$	   \\
53968.0083	&	0.2303	&	23.4	&	$-4.6$	&	331.1	    &	2.6	       \\
53968.0156	&	0.2426	&	23.2	&	$-4.8$	&	333.0	    &	1.7	       \\
53968.0611	&	0.3195	&	26.6	&	$-1.6$	&	307.5	    &	$-5.6$	   \\
53968.0686	&	0.3322	&	27.9	&	$-0.8$	&	297.5	    &	$-7.0$	   \\
53970.0325	&	0.6527	&	89.4	&	2.1	    &	$-164.6$	&	$-4.7$	   \\
53970.0937	&	0.7561	&	95.9	&	5.4	    &	$-213.9$	&	0.1	       \\
53970.0990	&	0.7651	&	95.8	&	5.3	    &	$-212.9$	&	0.9	       \\
53970.9646	&	0.2286	&	23.5	&	$-4.6$	&	330.8	    &	2.7	       \\
53970.9888	&	0.2695	&	23.4	&	$-4.4$	&	331.2	    &	$-0.6$	   \\
53970.9935	&	0.2775	&	23.7	&	$-4.1$	&	329.2	    &	$-1.3$	   \\
53971.0383	&	0.3532	&	30.6	&	0.4	    &	277.6	    &	$-9.4$	   \\
53972.1021	&	0.1518	&	29.9	&	0.4	    &	282.9	    &	7.1	       \\
53972.1060	&	0.1584	&	29.0	&	$-0.2$	&	289.3	    &	6.8	       \\
53972.1294	&	0.1980	&	25.1	&	$-3.1$	&	318.8	    &	4.6	       \\
53972.1333	&	0.2046	&	24.6	&	$-3.5$	&	322.2	    &	4.3	       \\
53972.1368	&	0.2105	&	24.3	&	$-3.9$	&	324.9	    &	3.9	       \\
53972.1411	&	0.2178	&	23.9	&	$-4.2$	&	327.7	    &	3.4	       \\
53975.8757	&	0.5320	&	66.9	&	2.0	    &	4.6	        &	$-12.3$	   \\
53975.9477	&	0.6538	&	89.5	&	2.1	    &	$-165.7$	&	$-4.7$	   \\
53975.9816	&	0.7111	&	94.9	&	4.4	    &	$-206.1$	&	$-3.0$	   \\
53976.0075	&	0.7549	&	95.9	&	5.4	    &	$-214.0$	&	0.0	       \\
53976.0253	&	0.7850	&	95.1	&	4.7	    &	$-207.5$	&	2.7	       \\
53976.0511	&	0.8286	&	91.6	&	1.5	    &	$-181.3$	&	7.1	       \\
53976.0606	&	0.8447	&	89.7	&	0.0	   &	$-167.0$    &	8.5	       \\
\hline
    \end{tabular}
\end{center}
\end{table}

The {\sc korel} program, used in our study, applies Fourier transforms for spectral decomposition. Spectra obtained during eclipses (except the totalities) were not used for this, as complications, such as the Rossiter-McLaughlin effect \citep{Rossiter_1924, McLaughlin_1924}, come into play. The  {\sc korel} code does not address such contingencies.  Our RV measurements of the components of $\epsilon$ CrA from {\sc Korel} are presented in Table~\ref{tab:eps_cra_rvs} and displayed in Fig~\ref{fig:eps_cra_rvc}.

In this way, {\sc korel} led to the spectroscopic orbital parameters given in Table \ref{tab:korel_pars}, where symbols have their usual meanings and the uncertainties of parameters are calculated using a simplex procedure. Our velocity semi-amplitudes for the components turned out to be a few km/s higher than those listed in \cite{Rucinski_2020}, the mass ratio retaining a similar value ($q = 0.133$, see Table~\ref{tab:korel_pars}). Reconstructed primary and secondary spectra from order 103 ($\Delta\lambda\sim$5000-5550 \AA) are shown in the top panel of Fig.\ref{fig:epsCrA_disentangled}, where the lower panel shows the O--C residuals of the spectrum fittings.  Order 103 shows relatively strong Th I and Fe I features at $\lambda$5528 and $\lambda$5535, respectively. 


\begin{table}
\begin{center}
\caption{Comparison of {\sc korel} orbital parameters with those of \citet{Rucinski_2020} for $\epsilon$ CrA.
\label{tab:korel_pars}}
\begin{tabular}{lll}
\hline 
Parameter                     & This study                  & \citet{Rucinski_2020}         \\
\hline                                                                                    
$P$ (days)                    &   \multicolumn{2}{c}{0.59145447 (fixed)}                    \\
$T_0$ (HJD-2450000)           & 3966.6892 (0.0696)          &    8312.0716 (0.0004)         \\
$K_1$ (km s$^{-1}$)          & 36.4 (0.05)                 &    34.718 (0.084)             \\
$K_2$ (km s$^{-1}$)          & 273.7 (0.05)                &    267.13(1.37)               \\
$V_{\gamma}$ (km s$^{-1}$)          &   59.6 (0.05)               &      62.541(0.076)            \\
$a \sin\ i$ \:(R$_{\odot}$)      & 3.603(0.002)                &    3.527(0.016)               \\
$(M_1+M_2), \sin^3i$ \:(M$_{\odot}$)    & 1.828 (0.002)               &    1.685 (0.023)              \\
RMS (km/s)                    &    \multicolumn{1}{l}{0.22} &          --                   \\
\hline                                                                                      
\end{tabular}
\end{center}
\end{table}

\section{Photometry}
\label{sec:photometry}
\subsection{ S Ant Photometry } 
\label{sec:S_Ant_photometry}

\begin{figure}
\includegraphics[scale=0.41]{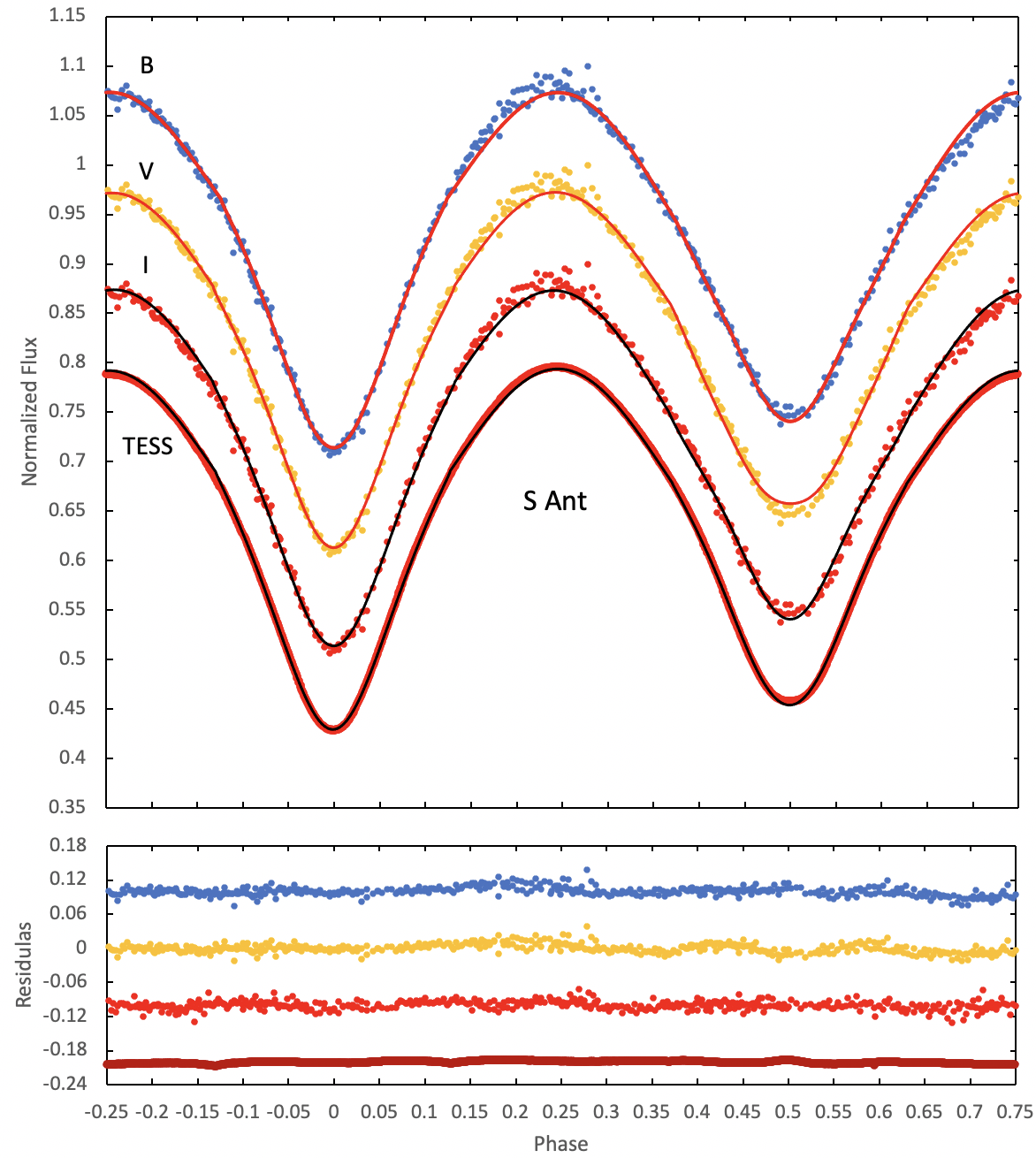}  
\caption{Ground-based BVI and TESS photometry of S Ant with the optimal model from {\sc WinFitter} fitting. Residuals to the model are plotted in the lower figure.
\label{fig:s_ant_all_lightcurve}}
\end{figure}
 
\begin{table}
\scriptsize
\caption{Optimal parameters for the {\sc WinFitter}  photometric model fits to the ground-based BVI and TESS Sector 9 data for S Ant. The conservative uncertainty estimates reflect the spread of numerous curve-fitting results, rather than the formal errors of particular fits.  
}
\label{tab:s_ant_lc_fitting}
\centering
\begin{tabular}{lllll}
\hline  
Parameter           &  B                  & V                 & I                   &  TESS Sector 9          \\ 
\hline 
$M_2/M_1$           & 0.35                & 0.35               & 0.35               & 0.34                    \\     
$L_1$               & 0.87 $\pm$ 0.09     & 0.82 $\pm$ 0.10    & 0.75 $\pm$ 0.01    &  0.71$\pm$ 0.02        \\  
$L_2$               & 0.13 $\pm$ 0.03     & 0.18 $\pm$ 0.05    & 0.25 $\pm$ 0.01    &  0.29 $\pm$ 0.01     \\  
$r_1 $ (mean)       & 0.47 $\pm$ 0.01     & 0.50 $\pm$ 0.02    & 0.47 $\pm$ 0.02    & 0.494 $\pm$ 0.005       \\ 
$r_2$ (mean)        & 0.27 $\pm$ 0.03     & 0.31 $\pm$ 0.05    & 0.31 $\pm$ 0.05    & 0.306 $\pm$ 0.013       \\ 
$i$ (deg)           & 75 $\pm$ 3          &  74 $\pm$ 2        & 72 $\pm$ 3         &  74.9 $\pm$ 0.5         \\ 
$T_h$ (K)           &      7100           &      7100          &       7100         &     7100                \\
$T_c$ (K)           &     6800            &      6800          &       6800         &     6800                \\     
$u_1$               & 0.60                & 0.50               & 0.32               & 0.29                    \\   
$u_2$               & 0.62                & 0.51               & 0.33               & 0.30                    \\ 
$\chi^2/\nu$        & 1.0                 & 1.0                & 1.0                & 0.77                    \\ 
$\Delta l $         & 0.01                & 0.01               & 0.01               & 0.005                   \\ 
\hline
\end{tabular}
\end{table}

{BVI} light curves of S Ant were collected by MB at the Congarinni Observatory (NSW 2447) using an 80mm f6 refractor, stopped down to 50mm to avoid saturation effects, and an Atik One 6.0 CCD camera equipped with Johnson-Cousins photometric filters. {\sc MaxIm DLTM} software was used for image calibration and aperture photometry. An ensemble of 3 comparison stars was used (TYC 6613 678, TYC 6613 1556, and TYC 6613 130). The adopted values for the reference were V = 6.725, B--V = 0.200 and V--I = 0.139. From this, we determined for S Ant: V = 6.295, B--V = 0.326, V--I = 0.196; V = 6.304, B--V - 0.326, V--I = 0.197 at the first and second maxima respectively. At the secondary occultation minimum, we found: V = 6.746, B--V = 0.325, V--I = 0.198. These measures are in keeping with the literature values cited in Section 1.1, but the colours appear a little more blue than typical Main Sequence values of type F3 \citep{Eker_etal_2018}.
 
The light curves are presented in  Fig~\ref{fig:s_ant_all_lightcurve}, together with the modelling results shown as smooth curves. We applied the optimal curve fitting program {\sc WinFitter} (WF, \cite{Rhodes_2023}) for the LC (light curve) modelling. The resulting parameter estimates are collected in Table~\ref{tab:s_ant_lc_fitting}.

LC modelling for close binary stars often refers to the numerical integration procedure of \citet{Wilson_1971} (hereafter referred to as WD), which represents the distorted component surfaces as equipotentials, according to the classical point-mass formulation attributed to \citet{Roche_1873} recalled in Ch.\ 3 of \citet{Kopal_1959}. Both the WD and WF methods converge to the same approximation for the surface perturbation when the internal structural constants $k_j$ are neglected, implying neglect also of the effects of tides on tides. The relevant formula,  Eqn 1-11 in Ch.1, or 2-6 in Ch. 3 of \citet{Kopal_1959}, is:
\begin{equation}
    \frac{ \Delta^{\prime} r }{r_0}  = q \sum^{4}_{j = 2} r_0^{j+1} (1 + 2k_j) P_j(\lambda) + n r_o^3(1 - \nu^2)  \,\,\, ,
    \label{eq:fosp}
\end{equation}
where $r$ is the local stellar radius expressed as a fraction of the orbital separation of the components with mean value $r_0$. The direction cosine of the angle between the radius vector $\hat{r}$ and the line of centres is here $\lambda$, and $\nu$ is the direction cosine of the angle between $\hat{r}$ and the rotation axis.  The coefficients $k_j$ (in WF)  can be taken from suitable stellar models, e.g.\ \citet{Inlek_2017}. They are set to zero in WD.

The surface perturbations ${\Delta}^{\prime} r$ are here serial harmonic functions, that start with terms of order $r^3$. The inclusion of terms of order $r^6$ and higher would imply the gravitational interaction of tides on tides being taken into account for self-consistency. Parallelism of WD and WF fitting functions, in which the mutual interactions of the  perturbations are neglected, therefore continues through to terms of order $r^5$, implying three additive tidal terms. Rotation with constant, synchronized, angular velocity, as usually assumed, and required in the Roche approximation, calls for only one source term. 

We can see from Table~\ref{tab:s_ant_lc_fitting}, and the contact criterion $r_1 + r_2 \approx 0.75$ (\citealt{Kopal_1959}; Table~3.3),  that the components of S Ant cannot be far from the idealized configuration of  Roche lobes osculating at the L$_1$ point. The mean radii approximate to values corresponding to the mass ratio $q \sim 0.3$ (see Table 3-1 of \citealt{Kopal_1959}). This prior value for $q$ tallies with a consensus of previous literature estimates and our own spectroscopic RVs discussed in Section~\ref{sec:S_Ant_spectrocopy}.  The WF posteriors are thus consistent with a near-Roche configuration.
   
We see directly from Fig~\ref{fig:s_ant_all_lightcurve} and Table~\ref{tab:s_ant_lc_fitting} that all four LCs show roughly the same ratio for the depths of the two minima ($\sim$0.8).  
The lack of variation in relative depths with colour implies the two stars must have close photospheric temperatures.  The light curves, by indicating a flatter central region of the secondary eclipse, point to S Ant being of the `A-type' \citet{Binnendijk_1970}, where the larger, slightly brighter, star eclipses its companion at the secondary minimum. In keeping with this lack of colour variation,  the ratio of $L_1/L_2$ is $\sim$4 in the {BVI}  wavebands.   This result aligns with a coarse estimate for a ratio of radii $\sim$0.6, in fair accord with the mass ratio, according to \citet{Kuiper_1941}'s approximation $r_2/r_1 \sim q^{0.46}$. In turn, this points to similar effective surface temperatures. 

This preliminary assessment thus raises the classical question for contact binaries \citep{Kuiper_1941}: why should two stars of such differing masses show closely similar surface temperatures?  Note that this point applies to the EW systems as a whole, and it bespeaks some special conditions that apply to the group generally.

\begin{table}[b]
\begin{center}
\caption{Optimal hot spot parameters for matching the O'Connell effect on S Ant. 
\label{tab:s_ant_oconnell_residuals_hot_spot} }
\begin{tabular}{llll}
    \hline
    Parameter       & B + V     & I            & TESS       \\
                    & (mean)    &              &  Sector 9  \\  
    \hline 
    $\lambda$ (deg)  & 72      &  82	       &  85.5      \\
    $\kappa$         & 1.29    &  1.28	       &  1.10      \\
    $U$              & 0.997   &  0.998	       &  0.998     \\
    \hline 
\end{tabular}
\end{center}
\end{table}      
               

\subsection{S Ant: O'Connell effect} 

\begin{figure}
    \centering
    \includegraphics[width=\columnwidth]{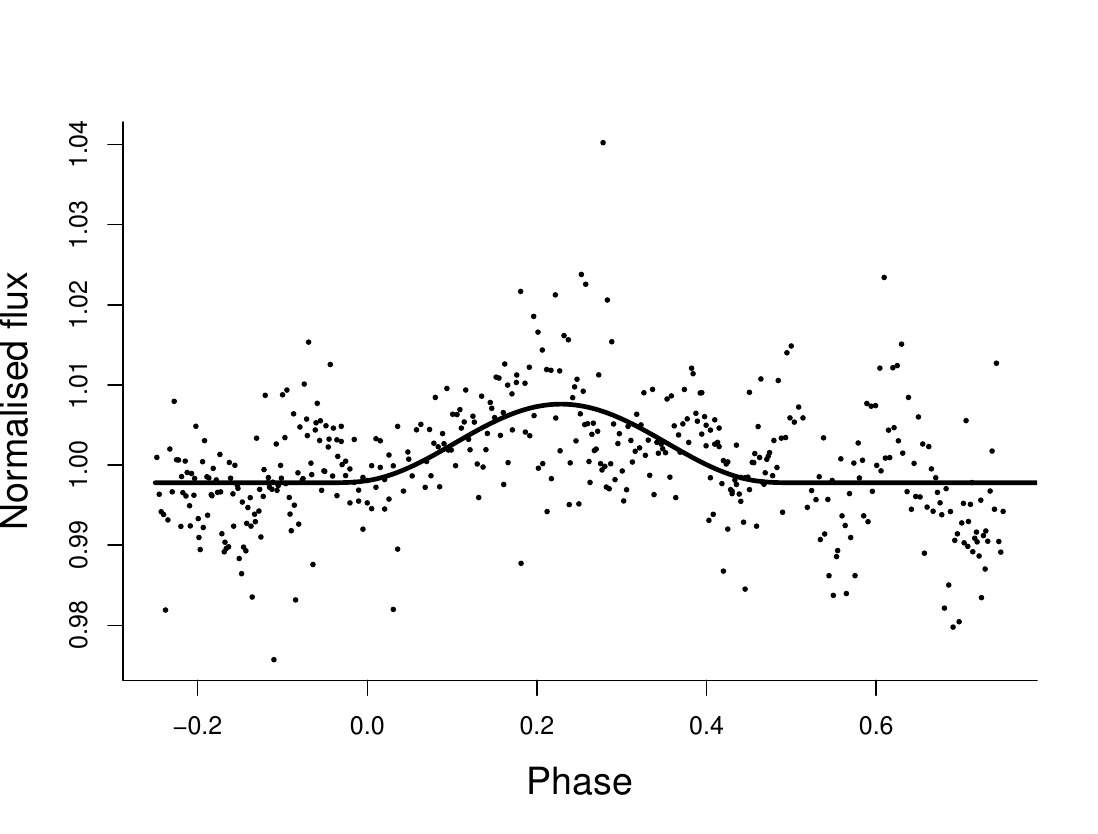}  
    \caption{Hot spot model for O'Connell effect in S Ant LC.
    \label{fig:s_ant_hot_spot_lightcurve}}
\end{figure}

The asymmetry in the LCs in Fig~\ref{fig:s_ant_all_lightcurve} --- in this case, the first maximum being higher than the second --- is easily seen. The TESS LCs (Sector 9) confirm the ground-based data on this point.  This is an example of the well-known  O'Connell effect \citep{Roberts_1906, OConnell_1951, Milone_1968}.   In what follows, we interpret this in terms of a locally heated region on the primary photosphere. This idea can be reconciled with other evidence.  Before following up on this, however, we should note that LC asymmetry is hardly present in classical photometric models. An initial LC fitting using a standard model can thus be expected to yield only approximate parameters. These may be improved in an iterative procedure if our interpretation of the spot's effects is adequate.

Fig~\ref{fig:s_ant_hot_spot_lightcurve} shows the fitting of a `hot spot' to the residuals from the standard model match, using the {\sc fitspot} application of the {\sc WinFitter} suite. {\sc fitspot} requires the specification of a circular spot's longitude $\lambda$, latitude $\beta$, angular size $\gamma$, inclination of the spin axis $i$, flux calibration unit $U$, the spot-bearing star's luminosity as a fraction of that of the whole system $L_1$, spot central relative flux, $\kappa$ and the limb-darkening coefficient $u_1$.  The appearance of Fig~\ref{fig:s_ant_hot_spot_lightcurve} --- i.e.\  the data scatter and simple form of the fitting function --- indicate the information content, or number of independently specifiable parameters, to be low. But we can reasonably adopt $\beta = 0$, $\gamma = 0.175$   ($\sim$10 deg), $i = 1.29$ (74 deg), $L_1 = 0.85$,  $u_1 = 0.5$. The remaining 3 parameters ($U, \lambda, \kappa $) were adjusted so as to minimize $\chi^2$. 

The flux excess of about 1.3 in the hot spot, in terms of the Planck function, corresponds to a temperature excess of about 500 K at 7000 K. In the IR, at a wavelength of about 1000 nm, this ratio drops to about 1.17.  We may therefore accept, referring to Table~\ref{tab:s_ant_oconnell_residuals_hot_spot}, that the O'Connell effect in these LCs of S Ant can be accounted for by a region of about 1.5{\%} of the primary's photospheric area heated to a temperature excess over the surrounding photosphere of several hundred kelvins. The implied energy release is $\sim 10^{32}$ erg s$^{-1}$.

This heating effect may be associated with the impact of a mass transferring stream, consistent with the general scenario of Roche Lobe overflow including the Coriolis deflection ( \citet{Lubow_1975}), its role in the semi-detached phase of the thermal relaxation oscillator (TRO) model \citep{Lucyb}, or the Algol-like version of this condition \citep{Stepien_2009}. In order to substantiate this picture, let us write the continuity of energy flow in the form
\begin{equation}
    L_{\rm spot} \approx\frac{1}{2} \dot{M} v^2 \,\,\, .
\label{eq:L_spot}
\end{equation}

In the following section, we will derive a value for $\dot{M}$ for the mass-losing star of at least  $\sim$10$^{18}$ gm  s$^{-1}$. The impacting kinetic term  $v^2/2$, per unit mass, in Eqn~\ref{eq:L_spot} comes from the drop in potential energy in moving from the region of the inner Lagrangian point to the impact location.  This can be written, approximately, as  $\sim \Delta \left( GM_1/r \right)$, where $M_1$ is the mass of the present primary, and $r$ is the radial separation from the primary centre of mass, relating to the stream's origin and impact positions. Eqn~\ref{eq:L_spot} will balance the luminosity of the spot with the dissipation of the stream's kinetic energy in the impact region with a value of the infall distance $\Delta r$ of not more than $\sim$10 \% of the primary mean radius. In this way, the O'Connell effect is explained consistently with the apparent secular change of period that will be discussed in section~\ref{s_ant_period}.


\subsection{\texorpdfstring{$\epsilon$ {CrA Photometry}} a }
\label{sec:eps_cra_photometry}

BVI photometry of $\epsilon$ CrA was carried out at the  Congarinni Observatory over 8 nights of reasonably stable weather in July and August 2021 using a 200mm f2.8 Canon lens on an SBIG STT3200 CCD camera.  The field of view was 4.38 x 2.95 deg, which enables the inclusion of a number of bright comparison stars. For data reduction, we used HIP 93825 ($V=4.209$, $B-V=0.522$, $V-I=0.417$) as the main comparison star.  Thence we determined for $\epsilon$ CrA: V = 4.766, B--V = 0.387, V--I = 0.260; V = 4.769, B--V - 0.376, V--I = 0.290; at the first and second maxima, respectively. At the occultation minimum, we found: V = 4.966, B--V = 0.403, V--I = 0.247. These measures are not far from the literature values cited in Section \ref{subsec:eps_cra_intro}, but the colours are a little more blue than the Main Sequence values of type F4 given by \citet{Eker_etal_2018}.

Photometric data from the Transiting Exoplanet Survey Satellite (TESS) was also used.  TESS collected data on $\epsilon$ CrA with a 2-min cadence in Sector 13 (June 20 to July 17, 2019).  These data are publicly accessible \citep{Ricker_etal_2015}. We downloaded and processed relevant information using the {\sc Lightkurve} Python package \citep{Lightkurve_2018}.

Photometric analysis was pursued using the numerical integration code of \citet{Wilson_1971} ({\sc WD}), combined with a Monte Carlo (MC) optimization procedure ({\bf see} \citep{Zola_2004}). This method models the LC of an eclipsing binary star, including proximity effects, adopting  Roche equipotentials \citep{Kopal_1959} for the component surfaces. The most important feature of the MC search procedure is its exploration of a grid of trial values for the input parameters (priors).  The method thus finds the model that provides a best fit to the observed LC by conducting up to hundreds of thousands of iterations, depending on the number of free parameters, with priors distributed over the selected input ranges, determined from preliminary expectations of component values and their expected uncertainties.

\begin{figure}
\centering
\includegraphics[scale=0.58]{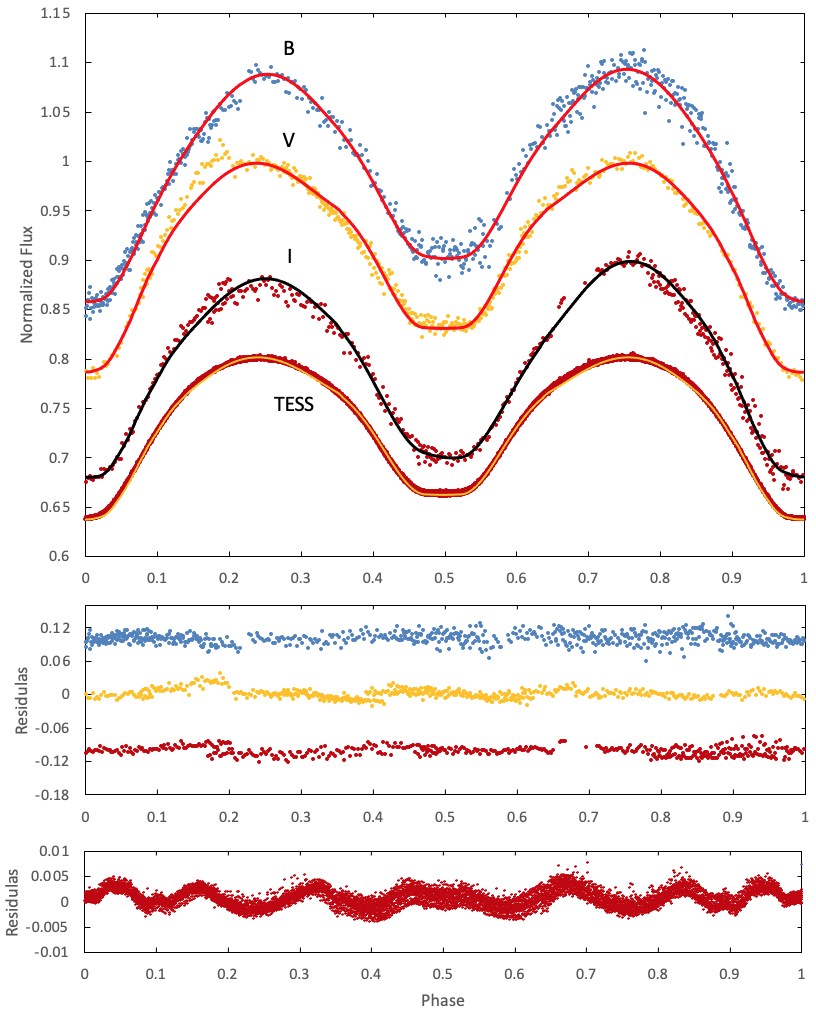}  
\caption{Ground-based BVI and TESS photometry of $\epsilon$ CrA with the optimal model from WD+MC (Wilson-Devinney plus Monte Carlo) fitting. Residuals to the model are plotted in the lower figures. 
\label{fig:bvi_lightcurve}}
\end{figure}

The BVI LCs of $\epsilon$ CrA show a slight O'Connell effect in the B  band, but the discontinuity in the residuals near the first maximum disturbs the V LC proportionately more than the B.
The asymmetry is more apparent in the I data, but the discontinuities seen in the residuals (Fig~\ref{fig:bvi_lightcurve}, lower panel) indicate that the asymmetries should be regarded cautiously. Note that there is very little  asymmetry between the maxima in the TESS LCs. 

Based on the previously cited literature 
we adopted the spectral type of $\epsilon$ CrA as F2V. Then, 
using the calibration data of both \citet{Pecaut_Mamajek_2013} and \citet{Eker_etal_2018}, we adopted the primary temperature 
as 6820 $\pm200$ K. 
The temperature of the secondary  ($T_2$) was adjustable on the range 5000 to 7000 K.
The input range of the orbital inclination was set to $50^\circ < i < 90^\circ$, 
The mass ratio ($q$) was fixed at the value of 0.13 obtained in the RV analysis in Section \ref{sec:eps_cra_spectroscopy}. 
The non-dimensional surface potentials 
were allowed to vary on the range 1.0 -- 3.0. 

Reasons for the asymmetry in the LCs include either hot spots that may relate to mass transfer or cool maculation.  The latter option was pursued for $\epsilon$ CrA on an empirical basis, and the results are presented in Table~\ref{tab:eps_cra_lc_fitting}.

For the {\sc WD+MC} program, the input range of $30^\circ < \beta < -30^\circ$ was set for the spot latitude and $0^\circ < \lambda < 360^\circ$ for the longitude. The input range of $10^\circ < \gamma < 50^\circ$ was set for the angular radius and $0.80 < \kappa < 1.3$ for the temperature factor.
Two approaches were used in dealing with the asymmetric LCs.  In the first approach, the separate bands and nights were treated on an individual basis, but this did not produce a uniform representation. A second approach adopted the TESS data as definitive, the asymmetry being very small. 
Two main spots were assumed to account for the B and I band asymmetries, while spots were not used in matching the almost symmetric V LC. In these fittings,  the geometric parameters were kept constant at the values from the TESS data analysis, while other parameters were adjusted to optimize the fitting.

Results are presented graphically in Fig.~\ref{fig:bvi_lightcurve}. 
A very small amplitude variation ($\sim$0.003 in relative flux) remains in the TESS LC residuals. Such residual effects notwithstanding, it is the TESS LC fitting results that are adopted in the absolute parameter calculations in (Section~\ref{sec:absolute_parameters}). The uncertainties given in Table~\ref{tab:eps_cra_lc_fitting}, that was subsequently used for calculating the formal error estimates of the absolute parameters in Table~\ref{tab:absolute_parameters}, were obtained from the {\sc WD+MC} code set with a 90\% confidence level.


\begin{table*}[t]
\caption{Optimal parameters for the WD+MC photometric model fits to the ground-based {BVI} and TESS Sector 13 data for $\epsilon$ CrA. According to the WD procedure for overcontact binaries, such as W UMa stars, these constraints were applied: the surface potential ($\Omega$), gravity brightening ($g$), bolometric albedo ($A$) and limb darkening parameters of components are the same. Here $g_1=g_2=0.32$ and $A_1=A_2=0.5$ are adopted. A quadratic limb-darkening law was applied with limb-darkening coefficients ($x$, $y$) taken from Claret (2017).
}
\label{tab:eps_cra_lc_fitting}
\centering
\begin{tabular}{lllll}
\hline  
Parameter       & B                 & V                 & I                 &  TESS Sector 13       \\ 
\hline 
$M_2/M_1$       & 0.13              & 0.13              & 0.13              &  0.13                 \\     
$L_1$           & 0.73 $\pm$ 0.02   & 0.82 $\pm$ 0.02   & 0.78 $\pm$ 0.02   & 0.65 $\pm$ 0.01       \\  
$L_2$           & 0.07 $\pm$ 0.01   & 0.10 $\pm$ 0.01   & 0.05 $\pm$ 0.01   & 0.09 $\pm$ 0.01       \\  
$L_3$           & 0.20 $\pm$ 0.01   & 0.08 $\pm$ 0.01   & 0.17 $\pm$ 0.01   & 0.26 $\pm$ 0.01       \\
$r_1 $ (mean)   & 0.56              & 0.56              & 0.56              & 0.56 $\pm$ 0.02       \\ 
$r_2$ (mean)    & 0.22              & 0.22              & 0.22              & 0.22 $\pm$ 0.01       \\ 
$i$ (deg)       & 74.02             & 74.02             & 74.02             & 74.02 $\pm$ 0.14      \\ 
$T_h$ (K)       & 6820              & 6820              & 6820              & 6820                  \\
$T_c$ (K)       & 6077 $\pm$ 157    & 6217 $\pm$ 117    & 5369 $\pm$ 143    & 6291 $\pm$ 109        \\     
$x$, $y$        & 0.233, 0.632      & 0.088, 0.689      & $-0.058$, 0.691   & 0.308, 0.223          \\   
Spot-1 Parameters   &               &                   &                   &                       \\
$\beta$ (deg)   & 74 $\pm$ 4        & --                & 31 $\pm$ 4        & --                    \\
$\lambda$ (deg) & 357 $\pm$ 10      & --                & 333 $\pm$ 12      & --                    \\
$\gamma$ (deg)  & 19 $\pm$ 2        & --                & 18 $\pm$ 2        & --                    \\
$\kappa$        & 0.78 $\pm$ 0.05   & --                & 0.55 $\pm$ 0.02   & --                    \\
Spot-2 Parameters   &               &                   &                   &                       \\
$\beta$ (deg)   & $-5 \pm$ 9        & --                & $-29 \pm$ 5       & --                    \\
$\lambda$ (deg) & 180 $\pm$ 5       & --                & 171 $\pm$ 11      & --                    \\
$\gamma$ (deg)  & 16 $\pm$ 2        & --                & 25 $\pm$ 5        & --                    \\
$\kappa$        & 0.60 $\pm$ 0.09   & --                & 0.52 $\pm$ 0.02   & --                    \\
\hline 
$\chi^2/\nu$    & 1.04              & 0.70              & 0.69              & 3.61                  \\ 
$\Delta l $     & 0.01              & 0.01              & 0.01              & 0.001                 \\ 
\hline
\end{tabular}
\end{table*}

\section{Orbital Period}
\label{sec:period}

\subsection{S Ant}
\label{s_ant_period}

\begin{figure}
  \centering
  \includegraphics[scale=0.48]{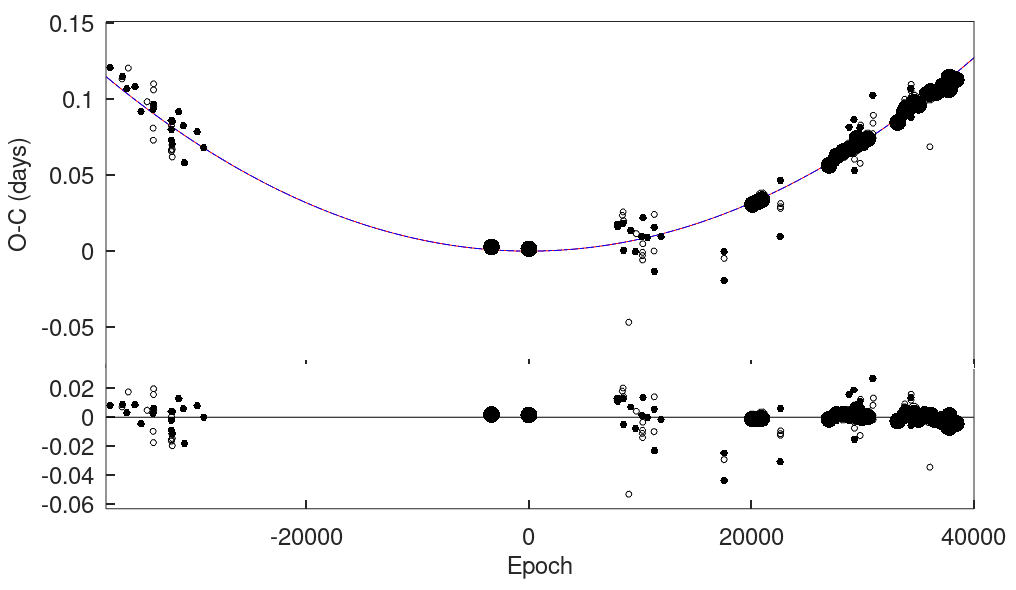}
   \caption{Observed minus calculated (O -- C) times of minimum light (ToMs)  for S Ant, and parabolic model fitting. Black points represent primary minima; blue open circles represent secondary minima. The sizes of the symbols are proportional to their weights. Residuals are shown in the lower panel.
   \label{S_Ant_O-C_diag}}
\end{figure}

Times of minimum light (ToMs) have been compiled from various sources, especially the O -- C {\em Atlas} data of \citet{Kreiner_2001} and \citet{Kreiner_2004}.  Observed ToMs are compared with calculated ones on the basis of an adopted reference ephemeris that is linear in the ToM number or epoch  $E$.  The corresponding time differences $T$ are used to form the familiar `O -- C' diagram (Fig.~\ref{S_Ant_O-C_diag}). For additional background see Ch.~8 of \citet{Budding_Demircan_2022}.  The ToMs used in  Fig.~\ref{S_Ant_O-C_diag} are listed in Table \ref{table:Min_S_Ant} in the Appendix. O--C data are matched with a trial function in $E$ to check for systematic departures from the adopted linear ephemeris. A parabola was fitted to the residuals for S Ant using {\sc Excel}$^{\rm TM}$. The results are given in Table~\ref{table:S_Ant_o_c}, and displayed in Fig.~\ref{S_Ant_O-C_diag}.

It can be shown that a steady transfer of mass between the stars in a close binary system will produce such a parabolic trend 
i.e.\ 
\begin{equation}
    T(E) = A + BE + C E^2  .
\label{par_model}
\end{equation}
The corresponding  rate of mass transfer takes the form:
\begin{equation}
    \dot{M_1} = 243.3 \frac{M_1 C}{g(x) P^2}   ,
\label{mass_transfer}
\end{equation}
where $M_1$ is the mass of the donor star, $P$ the orbital period. The function $g$ relates to the system's angular momentum and depends on the relative mass $ x = M_1/(M_1 + M_2)$. If we conserve the orbital angular momentum and neglect the contribution of the relatively small rotational terms, the function $g(x) = (2x - 1)/(1-x)$. 

The resulting value of $\dot{M_1}$ thus remains negative  for an upturned parabola ($C>0$) with $x< 1/2$ --- see Fig.~\ref{S_Ant_O-C_diag}, upper panel.  The residuals, after including the parabolic trend, are shown in the lower panel of Fig.~\ref{S_Ant_O-C_diag},   No further matching of residuals was attempted, and the parabola accounting for systematic effects to the available accuracy of the data. Following the foregoing prescription, our estimate for the rate of mass loss, in solar masses per year, is given in Table~\ref{table:S_Ant_o_c}.

\begin{table}[b]
    \begin{center}
        \caption{Parameters derived from O -- C analysis of S Ant.}
        \label{table:S_Ant_o_c}
        \begin{tabular}{ll}
            \hline 
            Parameter                               & Parabolic Model                                               \\
            \hline                                                                
            $T_0$ HJD                               & 2435139.9275 $\pm$ 0.0019                                     \\
            $P_{\rm orb}$ (d)                       & 0.64834512 $\pm$ 0.00000030                                   \\
            $C$ (d)                                 & 7.96$\times$ 10$^{-11}$ $\pm$ $1.92 \times 10^{-12}$          \\
            $\dot{M}$  (M$_{\odot}$ yr$^{-1}$)      & 2$\times 10^{-8}$                                             \\
            \hline                                                                
        \end{tabular}
    \end{center}
\end{table}

\subsection{\protect{$\epsilon$} CrA}
\label{eps_cra_period}

The orbital period of $\epsilon$ CrA was examined along the same lines as  S Ant. In this study,  4 primary and 4 secondary ToMs were obtained from the TESS data, while 3 primary and 3 secondary ToMs were obtained from the previously discussed BVI observations. These times of minima were combined with 23 primary and 8 secondary ToMs taken from \citet{Kreiner_2004}. All ToMs used in our analysis are given in Table~\ref{table:Min_eps_cra} in the Appendix.

\begin{figure}[t]
\centering
\includegraphics[scale=0.4]{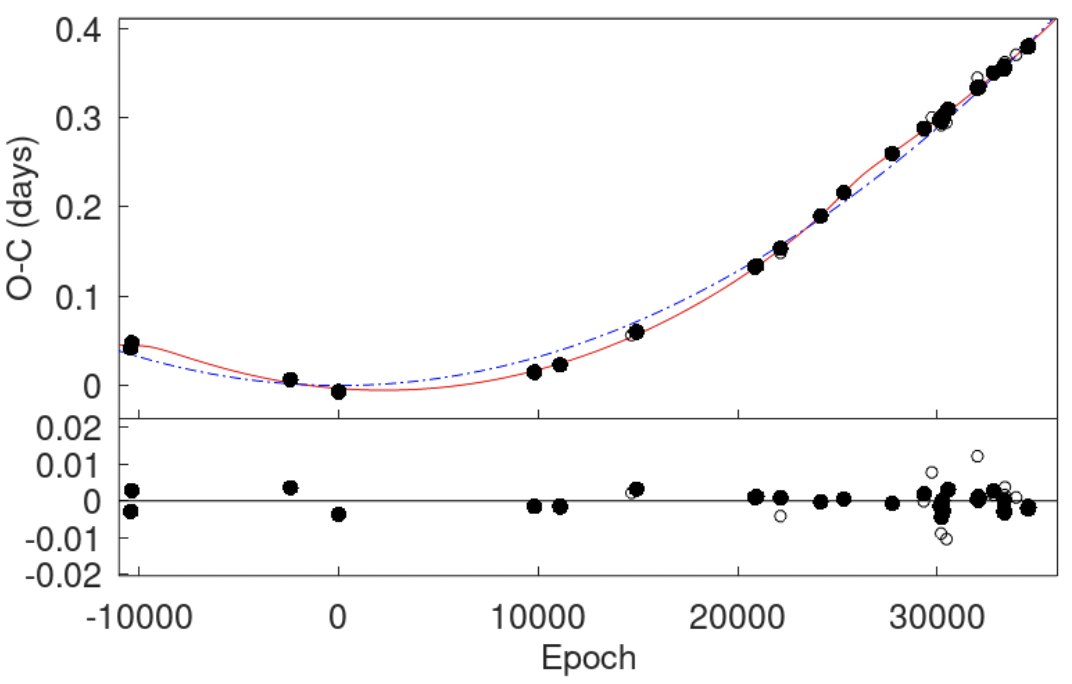} \\
\includegraphics[scale=0.4]{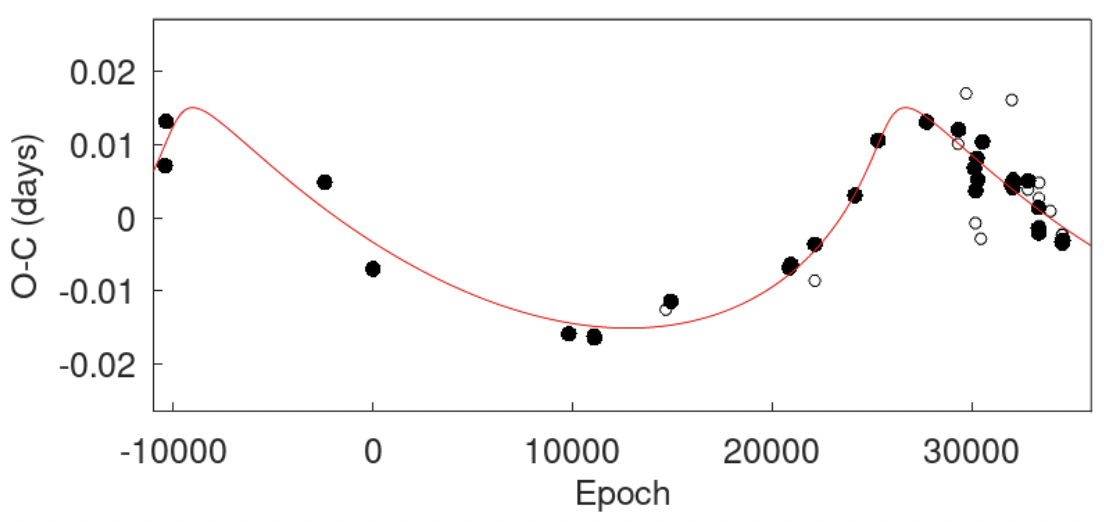} \\
\caption{LTE representation superimposed on parabolic form of O -- C changes of $\epsilon$ CrA (upper figure). Only LTE representation of O -- C changes after subtraction from the fitted parabola's residuals (lower figure). 
\label{fig:Eps_CrA_O-C_LTE}}
\end{figure}

The O -- C data using the linear ephemeris from \citet{Kreiner_2004} are shown in Fig.~\ref{fig:Eps_CrA_O-C_LTE}, which shows a relatively strong parabolic component. The corresponding form in Eq. \ref{par_model} was fitted to our data by the least squares method. Results are given in Table \ref{table:eps_cra_o_c}. The residuals from this fitting, shown in the lower panel, suggest a further, low-amplitude systematic effect.

This quasi-sinusoidal change in the O -- C diagram is suggestive of a light-time effect (LTE) associated with a third body.  In order to look into this, the first set of residuals was checked against the LTE formula given by \citet{Irwin_1959}:

\begin{multline}
    T(E) = A + BE + C E^2 \\ 
    + \frac{a_{12}\sin{i_3}}{c} \left( \frac{1-e_3^2}{1+e_3\cos{\nu_3}}\sin(\nu_{3} + \omega_{3}) + e_{3}\cos{\omega_3} \right),
\label{omega_obs}
\end{multline}
\noindent
where $c$ is the speed of light, and the other symbols have their usual meanings, as given by \citet{Irwin_1959}.
A weighted least-squares solution for $T_0$, $P_{orb}$, $Q$, $a_{12}$, $i_3$, $e_3$, $\omega_3$, $T_3$ and $P_{12}$ is presented in Table \ref{table:eps_cra_o_c}. The observational points and theoretical best-fit curve, and also the residuals, are plotted against epoch number and observation years in the lower part of Fig. \ref{fig:Eps_CrA_O-C_LTE}.


\begin{table}[b]
\centering
\scriptsize
\caption{Parameters derived from O -- C analysis of $\epsilon$ CrA.}
\label{table:eps_cra_o_c}
\begin{tabular}{lll}

\hline 
Parameter                   & Parabolic Model                                                   & LTE Model                                                 \\
\hline                                                                
$T_0$ HJD                   & 2439707.6661 $\pm$ 0.0024                                         & 2439707.6689 $\pm$ 0.0028                                 \\
$P_{orb}$ (d)               & 0.5914312 $\pm$ 0.0000002                                         & 0.5914318 $\pm$ 0.0000002                                 \\
$C$ (d)                     & $3.46 \times 10^{-10}$ $\pm$ $6.67 \times 10^{-12}$               & $3.21 \times 10^{-10}$ $\pm$ $8.05 \times 10^{-12}$       \\
$a_{12}\sin{i_3}$ (AU)      &                                                                   & 2.94 $\pm$ 0.23                                           \\
$e_3$                       &                                                                   & 0.74 $\pm$ 0.08                                           \\
$\omega_3$ (deg)            &                                                                   & 51 $\pm$ 12                                               \\
$T_3$ (HJD)                 &                                                                   & 2391635.267 $\pm$ 0.011                                   \\
$P_{12}$ (yr)               &                                                                   & 57.83 $\pm$ 0.34                                          \\
$\sum (O-C)^2$              & 0.02328                                                           & 0.00599                                                   \\
\hline                                                             \end{tabular}
\end{table}

The low value of the projected separation of the close binary from the centre of mass ($a_{12} \sin i $) compared with the period and the known mass of the close binary stars indicate that the wide orbit axis is relatively close to the line of sight for third body masses comparable to that of the close pair. With a separation of order an arcsecond, such a companion star (that has to be more massive than $\sim$0.4 M$_{\odot}$) should be detectable astrometrically.

Alternatively, vagaries from the parabolic trend in the upper part of Fig~\ref{fig:Eps_CrA_O-C_LTE} may be reflecting short-term changes in the O'Connell asymmetry, that may have physical implications on flow or accretion irregularities as well as discrepancies in the timings of light minima (see lower sub-diagram in Fig~\ref{fig:Eps_CrA_O-C_LTE}).  Disparities in implied mass transfer rates associated with contact binaries are not uncommon  \citet{Qian_2001}.

\section{Absolute parameters}
\label{sec:absolute_parameters}

The classical `eclipse method' of deriving stellar absolute parameters involves combining the results of photometric and spectroscopic analyses. The results are given in Table~\ref{tab:absolute_parameters}.
In these calculations, the masses ($m$) of the components are derived from Kepler's third law using the orbital period ($P$) and inclination ($i$) determined from photometry, and the projected orbital separation ($a \sin i$) with the mass ratio ($q$) found from fitting the RV curves. The absolute radii ($R$) of the components then follow, since $R=r  a$, with the fractional radii ($r$) and inclination ($i$) obtained from the LC fittings. The surface gravities ($g$) were computed as $g=g_{\odot}  m / R^2$. The bolometric magnitudes ($M_{bol}$) and luminosities ($L$)  were calculated using the following two equations with the absolute radii and effective temperatures listed in Table \ref{tab:absolute_parameters}:
\begin{equation}
M_{bol}=M_{bol,\odot} - 10\log (T/T_{\odot}) - 5\log (R/R_{\odot}),
\end{equation}
and 
\begin{equation}
    L/L_{\odot}=10^{0.4(M_{bol,\odot}-M_{bol})}.
    \end{equation}
    
The solar values, adopted by IAU 2015 Resolutions B2 and B3, were used here. Absolute visual magnitudes $M_V$ were derived from the bolometric corrections, $BC$ = $M_{bol}$ - $M_{V}$.  $BC$ values were taken from the online tabulation of \citet{Pecaut_Mamajek_2013} with the adopted effective temperatures.

The distance to the system was calculated using the absolute
magnitude formula: $M_{V}$ = $m_{V}$ --$A_{V}$ +  5 -- 5 log$(d)$.  
The Gaia parallaxes for S Ant and $\epsilon$ CrA point to interstellar absorption being of order 0.03 magnitudes and thus having a negligible effect on the photometric parallaxes.

The distance to S Ant and $\epsilon$ CrA 
were derived as $79 \pm$8 and $31 \pm$3 pc, which match the distance of $79 \pm$1 and $31 \pm$1 given by Gaia-DR3 \citep{Gaia_2022}, respectively.  \citet{Poro_etal_2024} recently found that the use of  Gaia DR3 parallaxes for 48 contact binary stars was successful in confirming the absolute parameters required for the corresponding photometric parallaxes. The photometric parallax formulae \citep{Popper_1998, Budding_Demircan_2007} produces distances of 85 $\pm10$ and 33 $\pm5$ pc for S Ant and $\epsilon$ CrA, thus supporting the reliability of our absolute parameter results.

\begin{table}
\begin{center}
\caption{Absolute parameters of S Ant and $\epsilon$ CrA.}
\label{tab:absolute_parameters}
\begin{tabular}{lll}
\hline
Parameter           & S Ant                 & $\epsilon$ CrA               \\
\hline 
$P$ (d)             & 0.64834976 $\pm$ 0.00000006 &  0.59145447 $\pm$ 0.0000003  \\
$i$ (deg)           & 76.58 $\pm$ 0.07	    & 74.02 $\pm$ 0.14             \\
$a$ ($\sun$)        & 4.10 $\pm$ 0.05       & 3.82 $\pm$ 0.02              \\
$m_1$ ($\sun$)      & 1.66 $\pm$ 0.10       & 1.89 $\pm$ 0.16              \\
$m_2$ ($\sun$)      & 0.55 $\pm$ 0.05       & 0.25 $\pm$ 0.04              \\
$R_1$ ($\sun$)      & 2.09 $\pm$ 0.11       & 2.14 $\pm$ 0.09              \\
$R_2$ ($\sun$)      & 1.31 $\pm$ 0.06       & 0.84 $\pm$ 0.04              \\
log $g_1$  			& 4.02 $\pm$ 0.03	    & 4.05 $\pm$ 0.03              \\
log $g_2$  			& 3.94 $\pm$ 0.02	    & 3.99 $\pm$ 0.03              \\
$T_1$ K	            & 7100 $\pm$ 200        & 6820 $\pm$ 200               \\
$T_2$ K	            & 6859	$\pm$ 200       & 6050 $\pm$ 100               \\
$L_1$ ($\sun$)      & 10.0	$\pm$ 2.2       & 8.92 $\pm$ 1.8               \\
$L_2$ ($\sun$)      & 3.4 $\pm$ 0.7         & 0.85 $\pm$ 0.14              \\
$M_{bol,1}$ (mag)   & 2.26 $\pm$ 0.24       & 2.38 $\pm$ 0.22              \\
$M_{bol,2}$ (mag)   & 3.42 $\pm$ 0.23       & 4.93 $\pm$ 0.20              \\
$M_{V,1}$ (mag)     & 2.25 $\pm$ 0.24       & 2.39 $\pm$ 0.22              \\
$M_{V,2}$ (mag)     & 3.42 $\pm$ 0.23       & 4.98 $\pm$ 0.20              \\
$V$ (mag)           & 6.45 $\pm$ 0.02       & 4.83 $\pm$ 0.02              \\
$d$ (pc)            & 79 $\pm$ 8            & 31 $\pm$ 3                   \\
\hline
\end{tabular}
\end{center}
\end{table}


\section{Discussion and conclusions}

\begin{figure*}
    \centering
    \includegraphics[scale=0.7]{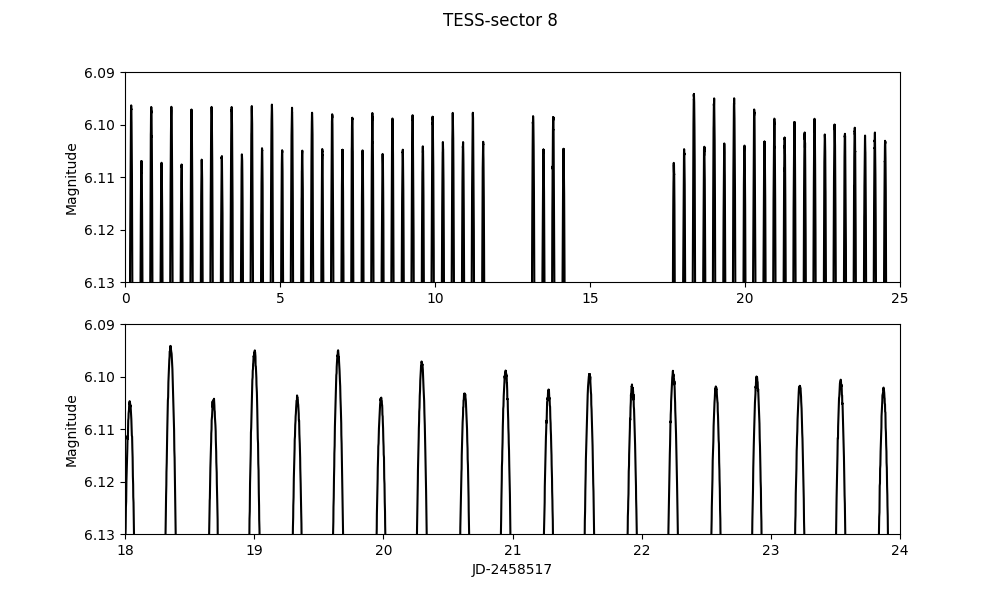} \\
    \caption{A succession of (partial, showing the maxima) TESS light curves showing short-term variability of the O'Connell effect. 
    \label{fig:TESS_OConnell}}
\end{figure*}

 
Ongoing discussion of the contact binary syndrome centres on the comparison of the various strands of observational evidence with theory -- such as the TRO scenario. Relaxation oscillations are characterized by a repetitive switching between two unstable end states. The behaviour in between is associated with charging and discharging phases, over time intervals connected with the Kelvin time scales of the components. 

In these terms, a young contact binary system might be observed in the discharging phase of the more massive primary, corresponding to a transfer of envelope material to the secondary.  In the original TRO scenario \citep{Lucyb}, the secondary then undergoes an energizing compression from a normal Main Sequence-like condensation, in order to comply with the equipotential layered structure of near-contact configurations. Secondary mass ejection with primary charging follows later, during a period of semi-detached separation from the common convective envelope. In due course, the secondary discharge subsides and the primary resumes what is effectively an interrupted Case A evolution process. This account addressed primarily the incidence of the cooler Binnendijk-type W contact binaries, taken to fall within a certain limited domain of $(J, q, M_1)_{\rm init}$ hyperspace characterising the W UMa type stars. Insufficient data on the A types allowed Lucy and others to surmise that these had evolved into a quasi-static thermal equilibrium condition.

In fact, for the bright  Binnendijk-class A binaries S Ant and $\epsilon$ CrA, evidence supports a semi-detached state with matter being now transferred from secondary to primary. Such binaries have a structural resemblance to low mass, short-period  Algols, with the proviso that the present primary is close to its own surrounding `Roche Lobe'. Statistical surveys, such as those of \citet{Hilditch_1988},  \citet{Maceroni_1996}, \citet{Drymova_1999}, \citet{Eker_2009};  have allowed the A systems to be seen as a continuation of the contact condition to higher masses and luminosities.  At higher mass we can expect the primary to have a correspondingly higher surface temperature. During the semi-detached stage, the transferred matter may form an equatorial belt around the primary where frictional effects detract from the orbital angular momentum \citep{Flannery, Stepien_2004, Stepien_2009}. 

The separation of the components depends on how the angular momentum is distributed, becoming sensitive to low values of the mass ratio.  The separation should decrease during discharges of the more massive star, pulling in the secondary and, if there is angular momentum loss from the orbit, tending ultimately towards merger. 
But continued mass loss from the original primary now increases the separation.  According to Henneco et al.\ (2024), faster, Case B, interaction can take place for mass ratios comparable to that of S Ant.  This suggests that the R CMa group of low angular momentum Algols (Budding, 1985) may have passed through a stage of similarity to S Ant at the present time. The  Algol-like, mass-transferring condition of S Ant is substantiated by its light curve asymmetry together with its consistent rate of period variation.   Fig~\ref{fig:TESS_OConnell} shows that the O'Connell effect has appreciable changes over relatively short timescales. 
Such variations can be regarded as flow or accretion structure instabilities, consistent with mass transfer.

The evolutionary path appears less definite with the relatively low mass ratio and angular momentum system $\epsilon$ CrA.  
While an Algol-like mass transfer process similar to that of S Ant could still be inferred, a future episode of primary discharge,  with the erstwhile primary now close to the Darwin stability limit, could result in a merger before the R CMa-like stage is attained.


In summary, this paper presents a detailed analysis 
of new high-quality data to yield reliable physical parameters of the W UMa type binaries S Ant and $\epsilon$ CrA.  This will help in the understanding of these complex systems.  Our results support the broadly Algol-like transfer of matter, 
consistent with the general framework of interactive binary evolution and concomitant effects.  However,  further questions are raised by this interpretation, which falls short of a full explanation of the contact binary syndrome, while still allowing alternative views. We offer this contribution as a pointer to further research on contact binary systems and related mass flow dynamics.

\section{Acknowledgments}
Generous allocations of time on the 1m McLennan Telescope and {\sc hercules} spectrograph at the Mt John University Observatory in support of the Southern Binaries Programme have been made available through its TAC and supported by its  Director, Dr.\ K.\ Pollard and previous Director, Prof.\ J.\ B.\ Hearnshaw. Useful help at the telescope was provided by the MJUO management (N.\ Frost and previously A.\ Gilmore \& P.\ Kilmartin). Considerable assistance with the use and development of the {\sc hrsp} software was given by its author Dr.\ J.\ Skuljan, and very helpful work with initial data reduction was carried out by R.\ J.\ Butland. VB would like to thank TUBITAK for its support within the scope of the 2214 overseas PhD research project program in 2006. Observations of $\epsilon$ CrA were made within the scope of this support. We thank the University of Queensland for the collaboration software. This paper includes data collected by the TESS mission and obtained from the MAST data archive at the Space Telescope Science Institute (STScI). STScI is operated by the Association of Universities for Research in Astronomy, Inc., under NASA contract NAS 5–26555.   This research has made use of the SIMBAD database, operated at CDS, Strasbourg, France, and of NASA's Astrophysics Data System Bibliographic Services. We thank the unnamed referee for  informative and helpful comments. The final version of this paper has improved significantly as a result.

\printbibliography
\onecolumn
\appendix
\section{Times of Minima}
\begin{longtable}{lllll}
\caption{Times of eclipses for S Ant. Column `type' indicates if the timing is for the primary eclipse (`pri') or the secondary (`sec').
Under `Filter/Source' the code `pg' refers to the timing being based on photographic data, `vis' visual, `pe' photo-electric, `BVI' to BVI photometry, and `R' to Johnson R photometry. `TESS' refers to the TESS satellite observing since 2018 in a special filter \protect\citep{Ricker_etal_2015}. `KWS' is the Kamogata/Kiso/Kyoto wide-field survey, observing in BVIc filters \protect\citep{Maehara_2014}. `OMC' are based on five-cm cameras onboard the INTEGRAL satellite, observing in the V filter since 2002 \protect\citep{Mas-Hesse_2003}.   `HIP' refers to the HIPPARCOS satellite, observing in a special Hp filter, between 1989 and 1993 \protect\citep{Perryman_1997}.  `ASAS' is the All Sky Automated Survey, observing since 1997 in the V and I filters \protect\citep{Pojmanski_2002}.  `CCD' is self-explanatory.  The column `Reference' otherwise indicates the source of the data. }
\label{table:Min_S_Ant}
\\
\hline \multicolumn{1}{l}{\textbf{HJD - 2400000}} & \multicolumn{1}{c}{\textbf{Error (d)}} & \multicolumn{1}{l}{\textbf{Type}} 
& \multicolumn{1}{l}{\textbf{Filter/Source}} & \multicolumn{1}{l}{\textbf{Reference}}\\ \hline 
\endfirsthead
\multicolumn{5}{c}%
{{\bfseries \tablename\ \thetable{} -- continued from previous page}} \\
\hline \multicolumn{1}{l}{\textbf{HJD - 2400000}} & \multicolumn{1}{c}{\textbf{Error (d)}} & \multicolumn{1}{l}{\textbf{Type}} 
& \multicolumn{1}{l}{\textbf{Filter/Source}} & \multicolumn{1}{l}{\textbf{Reference}}\\ \hline 
\endhead
\hline \multicolumn{5}{r}{{Continued on next page}} \\ \hline
\endfoot
\hline \hline
\endlastfoot
10741.5248  &               & pri       & pg         & \citet{Hogg_Bowe_1950}    \\
11435.5708	&               & sec 	    & vis	     & \citet{Luizet_1904}       \\
11460.5335	&               & pri 	    & vis	     & \citet{Luizet_1904}       \\
11715.9735	&               & pri 	    & vis	     &	\citet{Luizet_1904}      \\
11802.5413	&               & sec 	    & vis	     & \citet{Luizet_1904}       \\
12188.6185	& 	            & pri 	    & vis	     & \citet{Luizet_1904}       \\
12544.5437	&               & pri 	    & vis	     & \citet{Luizet_1904}       \\
12895.6290	&	        	&	 sec 	&	vis	     &	\citet{Yendell_1895}     \\
13247.6630	&	        	&	 sec 	&	vis	     &	\citet{Yendell_1895}     \\
13248.6480	&	        	&	 pri 	&	vis	     &	\citet{Yendell_1895}     \\
13262.5670	&	        	&	 sec 	&	vis	     &	\citet{Yendell_1895}     \\
13271.6770	&	        	&	 sec 	&	vis	     &	\citet{Sperra_1898}      \\
13274.5850	&	        	&	 pri 	&	vis	     &	\citet{Sperra_1898}      \\
13275.5710	&	        	&	 sec 	&	vis	     &	\citet{Sperra_1898}      \\
13286.5780	&	        	&	 sec 	&	vis	     &	\citet{Sperra_1898}      \\
14311.5830	&	        	&	 sec 	&	 pg 	 & \citet{Doberck_1899}      \\
14312.5620	&	        	&	 pri 	&	 pg 	 &	\citet{Doberck_1899}     \\
14314.5140	&	        	&	 pri 	&	 pg 	 &	\citet{Doberck_1899}     \\
14315.4720	&	        	&	 sec 	&	 pg 	 &	\citet{Doberck_1899}     \\
14316.4650	&	        	&	 pri 	&	 pg 	 &	\citet{Doberck_1899}     \\
14317.4340	&	        	&	 sec 	&	 pg 	 &	\citet{Doberck_1899}     \\
14346.5970	&	        	&	 sec 	&	 pg 	 &	\citet{Doberck_1899}     \\
14361.5010	&	        	&	 sec 	&	 pg 	 &	\citet{Doberck_1899}     \\
14363.4510	&	        	&	 sec 	&	 pg 	 &	\citet{Doberck_1899}     \\
14364.4270	&	        	&	 pri 	&	 pg 	 &	\citet{Doberck_1899}     \\
14366.3870	&	        	&	 pri 	&	 pg 	 &	\citet{Doberck_1899}     \\
14735.3016 	& 	        	&	 pri 	&	vis	     &	\citet{Luizet_1904}      \\
15008.8940	&	        	&	 pri 	&	vis	     &	\citet{Sperra_1900}      \\
15074.3525	&	        	&	 pri 	&	vis	     &	\citet{Luizet_1904}      \\
15815.4314	&	        	&	 pri 	&	vis	     &	\citet{Luizet_1904}      \\
16193.4061	&	        	&	 pri 	&	vis	     &	\citet{Luizet_1904}      \\
32962.1390	&	        	&	 pri 	&	pe	     &	\citet{Popper_1956}      \\
35139.9290	&	0.0010	    &	 pri 	&	pe	     &	\citet{Popper_1956}      \\
40289.4250	&	        	&	 sec 	&	vis	     &	\citet{Diethelm_Locher_1969} \\
40290.3990	&	        	&	 pri 	&	vis	     &	\citet{Diethelm_Locher_1969} \\
40314.3860	&	        	&	 pri 	&	vis      &	\citet{Diethelm_Locher_1969} \\
40589.6160	&	        	&	 sec 	&	vis	     &	\citet{Diethelm_Locher_1970a} \\
40629.4840	&	        	&	 pri 	&	vis	     &	\citet{Diethelm_Locher_1970a} \\
40630.4640	&	        	&	 sec 	&	vis	     &	\citet{Diethelm_Locher_1970b} \\
40655.4000	&	        	&	 pri 	&	vis	     &	\citet{Diethelm_Locher_1970b} \\
40658.3370	&	        	&	 sec 	&	vis	     &	\citet{Diethelm_Locher_1970b} \\
41023.3590	&	        	&	 sec 	&	vis	     &	\citet{Diethelm_Locher_1971a} \\
41070.3540	&	        	&	 pri 	&	vis	     &	\citet{Diethelm_Locher_1971b} \\
41350.4250	&	        	&	 pri 	&	vis	     &	 \citet{Locher_1972}	\\
41401.3320	&	        	&	 sec 	&	vis	     &	 \citet{Grough_1972}	\\
41728.4200	&	        	&	 pri 	&	vis	     &	 \citet{Diethelm_1973a}	\\
41753.3690	&	        	&	 sec 	&	vis	     &	 \citet{Diethelm_1973a}	\\
41764.3930	&	        	&	 sec 	&	vis	     &	 \citet{Diethelm_1973a}	\\
41766.3330	&	        	&	 sec 	&	vis	     &	 \citet{Diethelm_1973a}	\\
41777.3655	&	        	&	 sec 	&	vis	     &	 \citet{Diethelm_1973b}	\\
41789.3770	&	        	&	 pri 	&	vis	     &	 \citet{Diethelm_1973b}	\\
42052.5920	&	        	&	 pri 	&	vis	     &	 \citet{Locher_1974}	\\
42433.4860	&	        	&	 sec 	&	vis	     &	\citet{Carnevali_etal_1975}	\\
42445.4960	&	        	&	 pri 	&	vis	     &	 \citet{Diethelm_1975}	\\
42446.4770	&	        	&	 sec 	&	vis	     &	 \citet{Diethelm_1975}	\\
42458.4340	&	        	&	 pri 	&	vis	     &	 \citet{Diethelm_1975}	\\
42838.3870	&	        	&	 pri 	&	vis	     &	 \citet{Locher_1976}	\\
46514.4940	&	        	&	 pri 	&	vis	     &	\citet{Braune_Hubscher_1987} \\
46515.4620	&	        	&	 sec 	&	vis	     &	\citet{Braune_Hubscher_1987} \\
46516.4200	&	        	&	 pri 	&	vis	     &	\citet{Braune_Hubscher_1987} \\
48161.9700	&	0.0046	    &	pri	    &	HIP	     &	This study		\\
48162.2956	&	0.0036	    &	sec	    &	HIP	     &	This study		\\
48500.4080	&		        &	pri	    &	HIP	     &	This study		\\
48706.9097	&	0.0054	    &	sec	    &	HIP	     &	This study		\\
48707.2313	&	0.0043	    &	pri	    &	HIP	     &	This study		\\
49783.4600	&	0.0040	    &	 pri 	&	vis	     &	\citet{Martignoni_1995} \\
49797.4190	&	0.0050	    &	 sec 	&	vis	     &	\citet{Martignoni_1995} \\
49798.4090	&	        	&	 pri 	&	vis	     &	 \citet{Diethelm_1996}	\\
49799.3630	&	        	&	 sec 	&	vis	     &	 \citet{Diethelm_1996}	\\
49810.3880	&	        	&	 sec 	&	vis	     &	 \citet{Diethelm_1996}	\\
52627.7968	&		        &	pri	    &	pe	     &	O-C gateway\footnote{http://var2.astro.cz/ocgate/}		\\
52636.2270	&	        	&	 pri 	&	vis	     &	\citet{Nagai_2003} \\
53027.1850	&	        	&	 pri 	&	vis	     &	\citet{Nagai_2005} \\
53040.1460	&		        &	pri	    &	vis	     &	O-C gateway	\\
53071.9194	&	0.0037	    &	pri	    &	ASAS-3	 &	This study		\\
53072.2443	&	0.0045	    &	sec	    &	ASAS-3	 &	This study	\\
53458.9845	&	0.0053	    &	pri	    &	ASAS-3	 &	This study	\\
53459.3066	&	0.0045	    &	sec	    &	ASAS-3	 &	This study	\\
53798.0850	&	        	&	 pri 	&	vis	     &	\citet{Nagai_2007} \\
53813.9570	&	0.0040	    &	sec	    &	ASAS-3	 &	This study		\\
53814.2792	&	0.0035	    &	pri	    &	ASAS-3	 &	This study		\\
54073.2970	&	        	&	 sec 	&	vis	     &	\citet{Nagai_2007} \\
54074.2700	&	        	&	 pri 	&	vis	     &	\citet{Nagai_2007} \\
54076.2300	&	        	&	 pri 	&	vis	     &	\citet{Nagai_2007} \\
54097.2790	&	        	&	 sec 	&	vis	     &	\citet{Nagai_2007} \\
54099.2360	&	        	&	 sec 	&	vis	     &	\citet{Nagai_2007} \\
54110.2420	&	        	&	 sec 	&	vis	     &	\citet{Nagai_2008} \\
54111.2070	&	        	&	 pri 	&	vis	     &	\citet{Nagai_2008} \\
54114.1400	&	        	&	 sec 	&	vis	     &	\citet{Nagai_2008} \\
54201.0177	&	0.0084	    &	sec	    &	ASAS-3 	 &	This study		\\
54201.3438	&	0.0057	    &	pri	    &	ASAS-3   &	This study		\\
54253.2160	&	0.0029	    &	pri	    &	OMC	     &	This study		\\
54253.5412	&	0.0040	    &	sec	    &	OMC      &	This study		\\
54428.2760	&	        	&	 pri 	&	vis	     &	\citet{Nagai_2008} \\
54439.2850	&	        	&	 pri 	&	vis	     &	\citet{Nagai_2008} \\
54440.2470	&	        	&	 sec 	&	vis	     &	\citet{Nagai_2008} \\
54468.1510	&	        	&	 sec 	&	vis	     &	\citet{Nagai_2009} \\
54554.0448	&	0.0065	    &	pri	    &	ASAS-3	 &	This study		\\
54554.3710	&	0.0063	    &	sec	    &	ASAS-3	 &	This study		\\
54878.8685	&	0.0049	    &	pri	    &	ASAS-3   &	This study		\\
54879.1946	&	0.0057	    &	sec	    &	ASAS-3   &	This study		\\
55173.2460	&	        	&	 pri 	&	vis	     &	\citet{Nagai_2010} \\
55185.2150	&	        	&	 sec 	&	vis	     &	\citet{Nagai_2010} \\
55187.1670	&	        	&	 sec 	&	vis	     &	\citet{Nagai_2010} \\
55198.1790	&	        	&	 sec 	&	vis	     &	\citet{Nagai_2011} \\
55213.1060	&	        	&	 sec 	&	vis	     &	\citet{Nagai_2011} \\
56629.7362	&	0.0073	    &	sec	    &	KWS-V	 &	This study		\\
56630.0595	&	0.0068	    &	pri	    &	KWS-V	 &	This study		\\
57012.2720	&	        	&	 sec 	&	vis	     &	\citet{Nagai_2015} \\
57014.2140	&	        	&	 sec 	&	vis	     &	\citet{Nagai_2015} \\
57017.1340	&	        	&	 pri 	&	vis	     &	\citet{Nagai_2015} \\
57019.3989	&	0.0062	    &	sec	    &	KWS-Ic 	 &	This study		\\
57019.7220	&	0.0039	    &	pri	    &	KWS-Ic	 &	This study		\\
57040.1530	&	        	&	 sec 	&	vis	     &	\citet{Nagai_2016} \\
57128.3215	&	0.0053	    &	sec	    &	KWS-V	 &	This study		\\
57128.6465	&	0.0047	    &	pri	    &	KWS-V	 &	This study		\\
57378.2630	&	        	&	 pri 	&	vis	     &	\citet{Nagai_2016} \\
57391.2390	&	        	&	 pri 	&	vis	     &	\citet{Nagai_2017} \\
57392.2000	&	        	&	 sec 	&	vis	     &	\citet{Nagai_2017} \\
57406.1320	&	        	&	 pri 	&	vis	     &	\citet{Nagai_2017} \\
57407.1220	&	        	&	 sec 	&	vis	     &	\citet{Nagai_2017} \\
57413.2670	&	        	&	 pri 	&	vis	     &	\citet{Nagai_2017} \\
57414.2580	&	        	&	 sec 	&	vis	     &	\citet{Nagai_2017} \\
57415.2160	&	        	&	 pri 	&	vis	     &	\citet{Nagai_2017} \\
57430.1240	&	        	&	 pri 	&	vis	     &	\citet{Nagai_2017} \\
57566.2831	&	0.0081	    &	pri	&	Gaia DR3	 &	Gaia DR3		\\
57566.6100	&	0.0048	    &	sec	&	Gaia DR3	 &	Gaia DR3		\\
57743.2806	&	0.0048	    &	pri	&	KWS-Ic	     &	This study		\\
57743.6035	&	0.0041	    &	sec	&	KWS-Ic	     &	This study		\\
57841.5054	&	0.0058	    &	sec	&	KWS-V	     &	This study		\\
57841.8279	&	0.0041	    &	pri	&	KWS-V	     &	This study		\\
58197.1293	&	0.0017	    &	pri &	CCD	         &	\citet{Richards_etal_2019} \\
58476.2389	&	0.0049	    &	sec	&	KWS-Ic	     &	This study		\\
58476.5655	&	0.0045	    &	pri	&	KWS-Ic	     &	This study		\\
58504.0851	&	        	&	 sec 	&	CCD	     &	\citet{Nagai_2020} \\
58518.7087	&	0.0009	    &	pri	&	TESS	     &	This study \\
58519.0336	&	0.0008	    &	sec	&	TESS	     &	This study \\
58527.1375	&	0.0007	    &	pri	&	TESS	     &	This study \\
58527.4620	&	0.0001	    &	sec	&	TESS	     &	This study \\
58540.1046	&	0.0006	    &	pri	&	TESS	     &	This study \\
58540.4291	&	0.0010	    &	sec	&	TESS	     &	This study \\
58545.2914	&	0.0007	    &	pri	&	TESS	     &	This study \\
58545.6159	&	0.0010	    &	sec	&	TESS	     &	This study \\
58554.3679	&	0.0007	    &	pri	&	TESS	     &	This study \\
58554.6928	&	0.0009	    &	sec	&	TESS	     &	This study \\
58567.3351	&	0.0006	    &	pri	&	TESS	     &	This study \\
58567.6599	&	0.0010	    &	sec	&	TESS	     &	This study \\
58828.6173	&	0.0053	    &	pri	&	KWS-V	     &	This study		\\
58828.9412	&	0.0050	    &	sec	&	KWS-V	     &	This study	\\
59188.7757	&	0.0044	    &	sec	&	KWS-Ic	     &	This study	\\
59189.0993	&	0.0052	    &	pri	&	KWS-Ic	     &	This study	\\
59256.5296	&	0.0006	    &	pri	&	TESS	     &	This study		\\
59256.8543	&	0.0009	    &	sec	&	TESS	     &	This study		\\
59264.3100	&	0.0006	    &	pri	&	TESS	     &	This study		\\
59264.6346	&	0.0009	    &	sec	&	TESS	     &	This study		\\
59278.5734	&	0.0006	    &	pri	&	TESS	     &	This study		\\
59278.8982	&	0.0010	    &	sec	&	TESS	     &	This study		\\
59303.5370	&	0.0040	    &	 sec 	&	CCD	     &	\citet{Paschke_2021} \\
59322.9860	&	0.0022	    &	sec	&	BVI	         &	This study		\\
59325.9020	&	0.0025	    &  pri	&	BVI	         &	This study		\\
59330.4430	&	0.0030	    &	 pri 	&	CCD	     &	\citet{Paschke_2021} \\
59562.5461	&	0.0044	    &	pri	&	KWS-V	     &	This study		\\
59562.8732	&	0.0053	    &	sec	&	KWS-V	     &	This study	\\
59621.2253	&	0.0021	    &	sec	&	R	         &	This study	\\
59628.0373	&	0.0014	    &	pri	&	R	         &	This study	\\
59645.2135	&	0.0062	    &	sec	&	KWS-Ic	     &	This study	\\
59645.5344	&	0.0051	    &	pri	&	KWS-Ic	     &	This study	\\
59647.1595	&	0.0020	    &	sec	&	R	         &	This study	\\
60024.4980	&		        &	sec	&	CCD	         &	\citet{Paschke_2023}		\\
60038.4380	&		        &	pri	&	CCD	         &	\citet{Paschke_2023}	\\
\hline
\end{longtable}

\begin{longtable}{lllll}
\caption{Times of eclipses for $\epsilon$ CrA. Columns are as for Table~\protect\ref{table:Min_S_Ant}.   \label{table:Min_eps_cra}} \\
\hline \multicolumn{1}{l}{\textbf{HJD - 2400000}} & \multicolumn{1}{c}{\textbf{Error (d)}} & \multicolumn{1}{l}{\textbf{Type}} 
& \multicolumn{1}{l}{\textbf{Filter/Source}} & \multicolumn{1}{l}{\textbf{Reference}}\\ \hline 
\endfirsthead
\multicolumn{5}{c}%
{{\bfseries \tablename\ \thetable{} -- continued from previous page}} \\
\hline \multicolumn{1}{l}{\textbf{HJD - 2400000}} & \multicolumn{1}{c}{\textbf{Error (d)}} & \multicolumn{1}{l}{\textbf{Type}} 
& \multicolumn{1}{l}{\textbf{Filter/Source}} & \multicolumn{1}{l}{\textbf{Reference}}\\ \hline 
\endhead
\hline \multicolumn{5}{r}{{Continued on next page}} \\ \hline
\endfoot
\hline \hline
\endlastfoot
33547.3570	&		     & pri	   & pe	          &	\citet{Cousins_1950}	      \\	
33579.3000	&	         & pri 	   & pg 	      &	\citet{Cousins_1950}	      \\	
38279.3678	&            & pri     & pe           & \citet{Manfroid_etal_1987}	  \\	
39707.6619	&            & pri     & pe           & \citet{Tapia_1969}            \\	
45509.6300  &            & pri     & pe           & \citet{Manfroid_etal_1984}    \\
46266.6710  &            & pri     & pe           &	\citet{Manfroid_etal_1984}    \\
46273.7680  &            & pri     & pe           & \citet{Manfroid_etal_1984}    \\	
48385.5087	& 0.0009     & sec     & CCD          &	\citet{Kreiner_2004}          \\	
48536.6231  & 0.0013     & pri     & CCD          & \citet{Kreiner_2004}          \\	
52042.1121  &            & pri     & CCD          & \citet{Kreiner_2004}          \\	
52092.9768	&            & pri     & CCD          & \citet{Kreiner_2004}          \\	
52800.6393	& 0.0004     & sec     & CCD          & \citet{Kreiner_2004}          \\	
52802.1229	& 0.0002     & pri     & CCD          & \citet{Kreiner_2004}          \\	
53991.5288	& 0.0013     & pri     & CCD          & \citet{Ogloza_etal_2008}      \\	
54675.8413	& 0.0010     & pri     & CCD          & \citet{Kreiner_2004}          \\	
56112.4730	&            & pri     & CCD          & \citet{Paschke_2019}          \\	
57044.8911	& 0.0011     & sec     & CCD          & \citet{Kreiner_2004}          \\	
57057.0178	& 0.0006     & pri     & CCD          & \citet{Kreiner_2004}          \\	
57285.0270	& 0.0040     & sec     & CCD          & \citet{Richards_etal_2016}    \\
57527.2160	& 0.0030     & pri     & CCD          & \citet{Richards_etal_2017}    \\
57560.0340	& 0.0040     & sec     & CCD          & \citet{Richards_etal_2017}    \\
57568.0230	& 0.0020     & pri     & CCD          & \citet{Richards_etal_2017}    \\
57604.1060	& 0.0030     & pri     & CCD          & \citet{Richards_etal_2017}    \\	
57628.9440	& 0.0020     & pri     & CCD          & \citet{Richards_etal_2017}    \\	
57720.9066	& 0.0012     & sec     & CCD          & \citet{Kreiner_2004}          \\	
57768.5317	& 0.0005     & pri     & CCD          & \citet{Kreiner_2004}          \\	
58633.2284	& 0.0008     & pri     & CCD          & \citet{Kreiner_2004}		  \\	
58638.2674  & 0.0010     & sec     & CCD          & \citet{Kreiner_2004}		  \\	
58656.8869	&	0.0016	 &	pri	   & TESS         & This study					  \\	
58657.1823	&	0.0010	 &	sec	   & TESS         &	This study					  \\	
58667.2370	&	0.0011	 &	sec	   & TESS         &	This study					  \\	
58667.5323	&	0.0010	 &	pri	   & TESS         &	This study					  \\	
58669.8989	&	0.001	 &	pri	   & TESS         &	This study					  \\	
58670.1940	&	0.0011	 &	sec	   & TESS         &	This study					  \\	
58670.4904	&	0.0010	 &	pri	   & TESS         &	This study					  \\	
58680.5447	&	0.0009	 &	pri	   & TESS         &	This study					  \\	
58680.8413	&	0.0013	 &	sec	   & TESS         &	This study					  \\	
58681.4321	&	0.0012	 &	sec	   & TESS         &	This study				  	  \\	
58681.7277	&	0.0011	 &	pri	   & TESS         &	This study					  \\	
59117.0372	&	0.0029	 &	 pri   & cc           & \citet{Richards_etal_2021}    \\
59117.9232	&	0.0047	 &	 sec   & cc           & \citet{Richards_etal_2021}    \\	
59426.9521	&	0.0016	 &	pri    & BV	          &	This study					  \\	
59431.0951	&	0.0016	 &	pri	   & B	          &	This study					  \\	
59437.0062	&	0.0014	 &	pri	   & I	          &	This study					  \\	
59442.0383	&	0.0010	 &	sec	   & V	          &	This study					  \\	
59450.9122	&	0.0010	 &	sec	   & B	          &	This study				      \\	
59458.0037	&	0.0009	 &	sec	   & V	          &	This study				      \\	
59785.0795	&	0.0046	 &	sec	   & R	          &	This study					  \\	
60127.2313	&	0.0009	 &	pri	   & TESS	      &	This study					  \\	
60127.5280	&	0.001	 &	sec	   & TESS	      &	This study					  \\	
60139.3571	&	0.0012	 &	sec	   & TESS	      &	This study				      \\	
60139.6517	&	0.0009   &	pri	   & TESS	      &	This study					  \\	
60153.8469	&	0.001	 &	pri	   & TESS	      &	This study					  \\	
60154.1428	&	0.0016	 &	sec	   & TESS	      &	This study					  \\	
\hline
\end{longtable}

\newpage
\section{Spectroscopy}
\begin{table}
\begin{center}
    \begin{tabular}{|c|c|c|c|r|}
    \hline
    Image       & Date          & Phase     & Exp. time (s) &  S/N \\
    \hline
w9236030s	&	21/01/2021	&	0.370	&	600	&	75	\\
w9236032s	&	21/01/2021	&	0.391	&	600	&	70	\\
w9236034s	&	21/01/2021	&	0.412	&	900	&	90	\\
w9236036s	&	21/01/2021	&	0.429	&	900	&	90	\\
w9236038s	&	21/01/2021	&	0.449	&	900	&	80	\\
w9236040s	&	21/01/2021	&	0.471	&	900	&	90	\\
w9236042s	&	21/01/2021	&	0.491	&	900	&	60	\\
w9236044s	&	21/01/2021	&	0.512	&	900	&	70	\\
w9236046s	&	21/01/2021	&	0.531	&	900	&	67	\\
w9236048s	&	21/01/2021	&	0.551	&	900	&	80	\\
w9238025s	&	23/01/2021	&	0.546	&	900	&	45	\\
w9238027s	&	23/01/2021	&	0.565	&	900	&	60	\\
w9238029s	&	23/01/2021	&	0.583	&	900	&	60	\\
w9238031s	&	23/01/2021	&	0.605	&	900	&	55	\\
w9238033s	&	23/01/2021	&	0.623	&	900	&	82	\\
w9238035s	&	23/01/2021	&	0.644	&	900	&	35	\\
w9239016s	&	24/01/2021	&	0.863	&	900	&	70	\\
w9239018s	&	24/01/2021	&	0.884	&	900	&	65	\\
w9239020s	&	24/01/2021	&	0.906	&	900	&	70	\\
w9239022s	&	24/01/2021	&	0.943	&	900	&	75	\\
w9239024s	&	24/01/2021	&	0.965	&	900	&	70	\\
w9239026s	&	24/01/2021	&	0.988	&	900	&	65	\\
w9239029s	&	24/01/2021	&	0.110	&	900	&	70	\\
w9240019s	&	25/01/2021	&	0.425	&	900	&	80	\\
w9240021s	&	25/01/2021	&	0.468	&	900	&	80	\\
w9240023s	&	25/01/2021	&	0.488	&	900	&	75	\\
w9240025s	&	25/01/2021	&	0.510	&	900	&	55	\\
w9240027s	&	25/01/2021	&	0.530	&	900	&	50	\\
w9240030s	&	25/01/2021	&	0.588	&	900	&	70	\\
w9240032s	&	25/01/2021	&	0.606	&	900	&	70	\\
w9240034s	&	25/01/2021	&	0.627	&	900	&	75	\\
w9240036s	&	25/01/2021	&	0.646	&	900	&	80	\\
w9240038s	&	25/01/2021	&	0.665	&	900	&	80	\\
w9240040s	&	25/01/2021	&	0.685	&	900	&	80	\\
w9240042s	&	25/01/2021	&	0.699	&	900	&	50	\\
w9241016s	&	26/01/2021	&	0.964	&	900	&	65	\\
w9241018s	&	26/01/2021	&	0.983	&	900	&	50	\\
w9241020s	&	26/01/2021	&	0.013	&	1800	&	75	\\
w9241022s	&	26/01/2021	&	0.039	&	900	&	75	\\
w9241024s	&	26/01/2021	&	0.059	&	900	&	70	\\
w9241026s	&	26/01/2021	&	0.077	&	900	&	65	\\
w9241028s	&	26/01/2021	&	0.101	&	900	&	65	\\
w9241030s	&	26/01/2021	&	0.124	&	900	&	70	\\
w9241032s	&	26/01/2021	&	0.149	&	1200	&	85	\\
w9241036s	&	26/01/2021	&	0.197	&	900	&	75	\\
w9241038s	&	26/01/2021	&	0.222	&	900	&	70	\\
w9241040s	&	26/01/2021	&	0.245	&	900	&	70	\\
w9241042s	&	26/01/2021	&	0.266	&	900	&	80	\\
w9241044s	&	26/01/2021	&	0.289	&	900	&	80	\\
    \hline
    \end{tabular}
    \caption{Log of spectroscopic observations for S Ant. S/N refers to 5500 \AA.}
    \label{tab:SAnt_obs_log}
\end{center}
\end{table}

\begin{table}
\begin{center}
    \begin{tabular}{|c|c|c|c|r|}
    \hline
    Image       & Date          & Phase     & Exp. time (s) &  S/N \\
    \hline
    w3966049s   & 19/8/2006     & 0.489     & 416           & 130 \\
    w3966050s   & 19/8/2006     & 0.498     & 409           & 110 \\
    w3967003s   & 19/8/2006     & 0.542     & 329           & 98  \\
    w3968002s   & 20/8/2006     & 0.229     & 652           & 138 \\
    w3968003s   & 20/8/2006     & 0.242     & 489           & 136 \\
    w3968009s   & 20/8/2006     & 0.319     & 527           & 131 \\
    w3968010s   & 20/8/2006     & 0.331     & 625           & 143 \\
    w3968016s   & 20/8/2006     & 0.404     & 901           & 140 \\
    w3968018s   & 20/8/2006     & 0.424     & 749           & 131 \\
    w3970005s   & 22/8/2006     & 0.652     & 435           & 112 \\
    w3970006s   & 22/8/2006     & 0.662     & 392           & 100 \\
    w3970012s   & 22/8/2006     & 0.729     & 1200          & 76  \\
    w3970013s   & 22/8/2006     & 0.755     & 399           & 97  \\
    w3970014s   & 22/8/2006     & 0.764     & 412           & 98  \\
    w3970067s   & 23/8/2006     & 0.108     & 221           & 128 \\
    w3970068s   & 23/8/2006     & 0.113     & 196           & 124 \\
    w3970088s   & 23/8/2006     & 0.228     & 239           & 131 \\
    w3970094s   & 23/8/2006     & 0.269     & 283           & 151 \\
    w3970095s   & 23/8/2006     & 0.277     & 379           & 154 \\
    w3971010s   & 23/8/2006     & 0.352     & 208           & 147 \\
    w3972022s   & 24/8/2006     & 0.151     & 304           & 128 \\
    w3972023s   & 24/8/2006     & 0.158     & 272           & 129 \\
    w3972029s   & 24/8/2006     & 0.197     & 301           & 126 \\
    w3972030s   & 24/8/2006     & 0.204     & 277           & 150 \\
    w3972031s   & 24/8/2006     & 0.210     & 210           & 132 \\
    w3972033s   & 24/8/2006     & 0.217     & 259           & 134 \\
    w3975018s   & 28/8/2006     & 0.531     & 900           & 126 \\
    w3975020s   & 28/8/2006     & 0.653     & 288           & 98  \\
    w3975026s   & 28/8/2006     & 0.710     & 361           & 89  \\
    w3976002s   & 28/8/2006     & 0.754     & 253           & 85  \\
    w3976008s   & 28/8/2006     & 0.784     & 261           & 67  \\
    w3976014s   & 28/8/2006     & 0.828     & 576           & 84  \\
    w3976016s   & 28/8/2006     & 0.844     & 751           & 86  \\
    w3976017s   & 28/8/2006     & 0.860     & 311           & 96  \\
    w3976019s   & 28/8/2006     & 0.871     & 351           & 110 \\
    w3976020s   & 28/8/2006     & 0.884     & 337           & 84  \\
    \hline
    \end{tabular}
    \caption{Log of spectroscopic observations for $\epsilon$ CrA. S/N refers to 5500 \AA.}
    \label{tab:epsCrA_obs_log}
\end{center}
\end{table}

\begin{table}
\fontsize{9pt}{9pt}\selectfont
    \centering
     {\begin{tabular}{cl}
        \hline
        Spectral & Comment \\
        Order & \\ 
        \hline
         85& No strong feature.\\
         86& Blue wing of H$\alpha$ + telluric intrusions visible.\\
         87& Most of order taken up with H$\alpha$.\\
         88& Fe I 6456 doublet + telluric intrusions.\\
         89& Si I (6671.36 \AA) tentatively  identified for primary.\\
         90& Fe I (6436.334 \AA) p + telluric lines.\\
         91& A wide blend measured at 6249.0\AA (vacuum).\\
         92& Very broad feature measured at 6139.9 \AA, attributed to Fe I. \\
         93& V I 6111.6 \AA \, and  Fe I 6113.4 \AA \, in broad blend.\\
         94& No visible lines.\\
         95& Telluric lines only .\\
         96& Relatively strong Na I (5890 \& 5896 \AA) doublet with strong telluric lines. \\
         97& Very broad feature around 5861. Probably Fe I blend.\\
         98& Weak metallic lines.\\
         99& Mg I blended triplet around 5711 \AA. \\
         100& Blended metallic lines. \\
         101& Fe I 5615.652 \AA \, Ni I 5625.326 \AA \, identified. \\
         102& Central peak of Ca I doublet measured at 5603.9 \AA (vac). \\
         103& Relatively strong Ba I  (5555.484 \AA) and Mg II (5528.409 \AA) lines.\\
         104& Fe I lines at 54479.4, 5448.3, 5457.6 \AA \, (vac) visible.\\
         105& Very broad feature at 5407.8 \AA  \ (vac) visble. Probably mainly 
         Fe I blend. \\
         106& Broad neutral iron lines at 5366.0, 5372.8 and 5384.4 \AA (vac).\\
         107& Strong Fe I  lines at 5319.1, 5329.4 and 5341.1 \AA (vac).  Some apparent asymmetry in lines.\\
         108& Fe I at 5283.628\AA \, and Ca I at 5264.239 \AA \, identified. \\
         109& Fe I lines at 5198.8, 5126.798 \& 5227.192 \AA \, , Co I at 5235.188 \AA \, identified.\\
         110&  Mg I triplet, and Cr I blend at 5183.6 and 5183.4 \AA \, .  Strong but complex blend. \\
         111& Ni I,  Fe I blends at 5099.6 and 5139.3 \AA. \\
         112& Fe I at 5050.1, Ni I at 5080.5 \AA \,.\\
         113& Measurable primary Fe I lines at 5018.4, 5041.3 and 5051.3 \AA \,.\\
         114& Strong primary Fe I (4985.6 \AA) with blends at 4992.8 and 5003.7 \AA \, 
         (Ni I).\\
         115& Fe I + Ni I at 4933.2.  Other Fe I lines at 4939.5, 4957.4 and Sr I at 4967.9 \AA \,.\\
         116& Cr II line at 4891.6\AA; Fe I at 4891.4, 4910.3 and 4920.5 \AA \,.\\
         117& H$\beta$ 4861.4 \AA \,, dominates the order, with Fe I intrusions in red wing. \\
         118& Isolated Cr I line at 4824.4 \AA \, usable for both   primary and secondary RV displacements.\\
         119& Ti II at 4764.5 \AA \,.\\
         120& Mg I at 4730.0, Fe I at 4736.8 and Mn doublet at 4762  \AA \,.\\
         121& Fe I 4680.0, Ti II at 4708.7 and Ti I at 4715,3 \AA \,.\\
         122& Strong blend of Fe I and Cr I (4667.5 \AA) + Ti I 4535.6, Fe II 4635.3.\\
         123&  Featureless.  \\
         124& Strong line of Fe II (4583.8 \AA) with some asymmetry\\
         125& Line features at (vac)  4535.9, 4543.3, 4551.7, 4557.6 4564.7 \& 4574.1 \AA \,  include Ti, Fe ana Co species.\\   
         \hline
    \end{tabular} }
    \caption{Line absorptions seen on the S Ant (UCMJO) spectra at elongation.
    \label{tab:SAnt_spectrum}}
\end{table}

\begin{table}
    \centering
    {\begin{tabular}{cl}
        \hline
        Spectral & Comment \\
        Order & \\
        \hline
         85& Relatively strong Fe I (6678 \AA) line.\\
         86& Blue part of the H$\alpha$ line wing.\\
         87& Part of the broad H$\alpha$ line.\\
         88& Some relatively weak metallic lines accompanying to strong Fe I (6495 \AA) line.\\
         89& Relatively strong Th I (6646 \AA) line at the blue edge of the order and some weak metallic lines.\\
         90& Some relatively strong metallic lines with telluric lines.\\
         91& Four moderate-strength metallic lines in the middle of the order.\\
         92& Relatively weak metallic lines.\\
         93& Relatively strong Fe I (6103 \AA) line with weak lines.\\
         94& Weak metallic lines.\\
         95& Weak metallic and telluric lines.\\
         96& Relatively strong Na I (5890 \& 5896 \AA) doublet, weak metallic and telluric lines. \\
         97& Some blended metallic lines in the middle of the order.\\
         98& Weak metallic lines.\\
         99& Two relatively strong Fe I lines accompanying blended lines on the blue edge of the order.\\
         100& Relatively strong but blended metallic lines.\\
         101& Some blended lines accompanying two relatively strong metallic lines.\\
         102& Relatively strong Fe II (5588 \AA) line accompanying weak blended lines.\\
         103& Relatively strong Th I (5528 \AA) and Fe I (5535 \AA) lines.\\
         104& Several strong metallic lines.\\
         105& Relatively weak metallic lines, mostly blended.\\
         106& Blended metallic lines, some are relatively strong.\\
         107& Strong metallic lines, some are blended.\\
         108& Strong blended lines of Fe I.\\
         109& Some relatively strong metallic lines.\\
         110& The region of Mg I triplet.\\
         111& Relatively strong line of Fe I (5099 \AA) with some blended lines.\\
         112& Some blended metallic lines.\\
         113& Some blended metallic lines.\\
         114& Relatively strong line of Fe I (4984 \AA) with some weaker and blended metallic lines.\\
         115& Relatively strong but blended metallic lines.\\
         116& Blue part of the H$\beta$ line wing with some metallic lines.\\
         117& The region of H$\beta$ with some artefacts due to CCD.\\
         118& Relatively weak metallic lines with some artefacts due to CCD..\\
         119& Broad and blended metallic lines.\\
         120& Relatively weak and blended metallic lines.\\
         121& Relatively weak and blended metallic lines.\\
         122& Relatively strong line of Fe I (4668 \AA) together with weaker metallic lines.\\
         123& Weak and blended metallic lines.\\
         124& Relatively strong line of Fe I (4583 \AA) together with weaker metallic lines.\\
         125& Some relatively strong metallic lines.\\
         \hline
    \end{tabular}}
    \caption{Some features seen on $\epsilon$ CrA spectrum taken at UCMJO.}
    \label{tab:epscra_spectrum}
\end{table}


\begin{figure*}
    \centering %
    \begin{subfigure}{0.32\textwidth}
      \noindent\includegraphics[width=\linewidth]{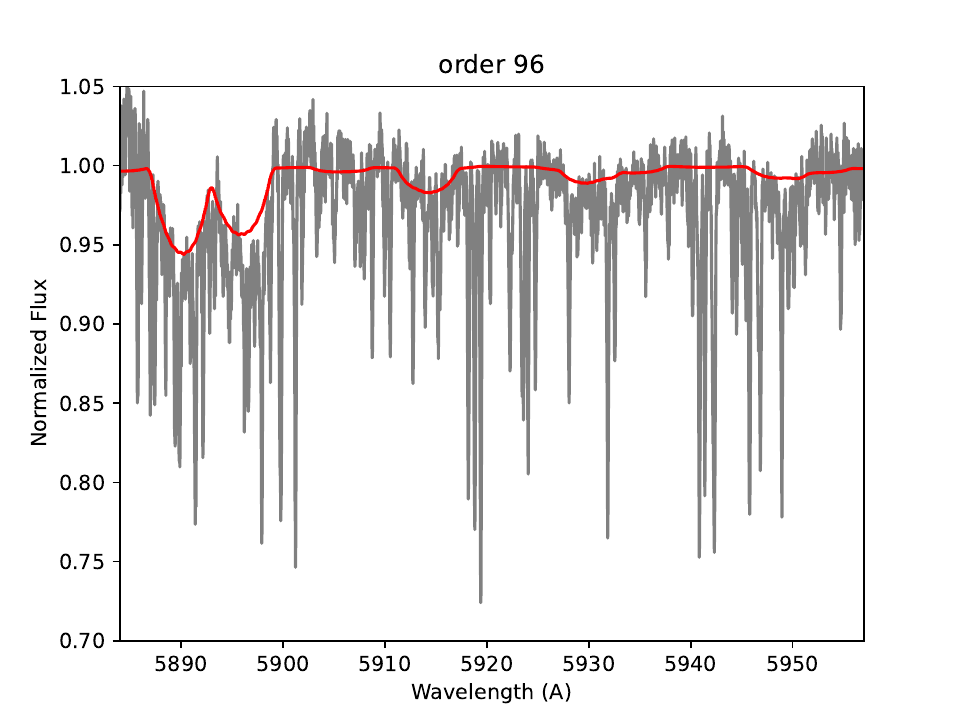}
        \caption{Order 96}
        \label{fig:s_ant_o96}
   \end{subfigure}
    \begin{subfigure}{0.32\textwidth}
        \includegraphics[width=\linewidth]{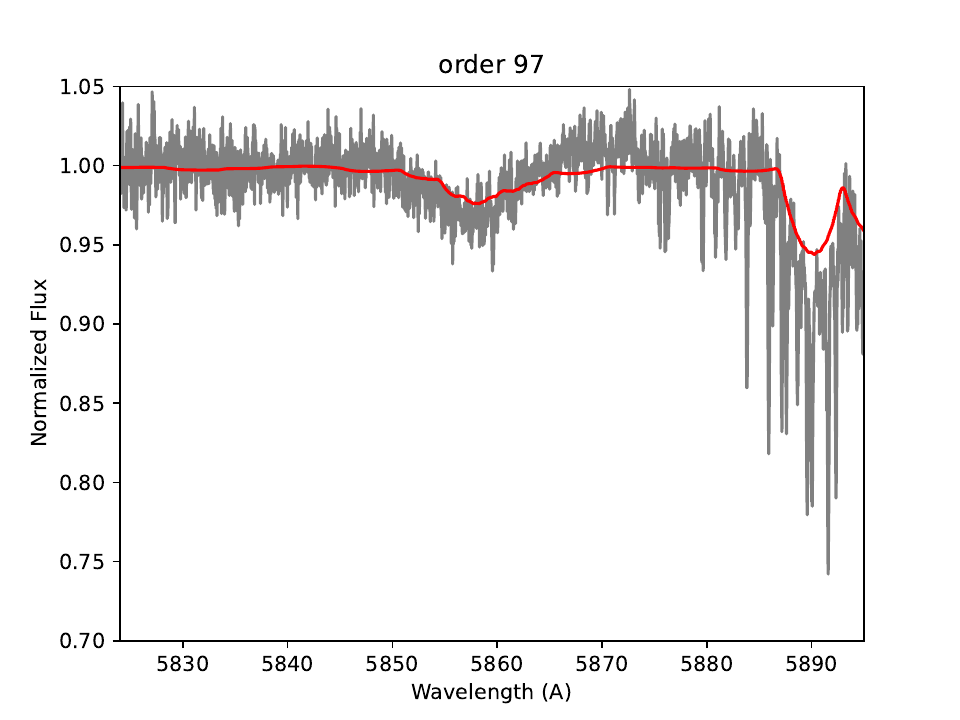}
        \caption{Order 97}
        \label{fig:s_ant_o97}
    \end{subfigure}
    \begin{subfigure}{0.32\textwidth}
        \includegraphics[width=\linewidth]{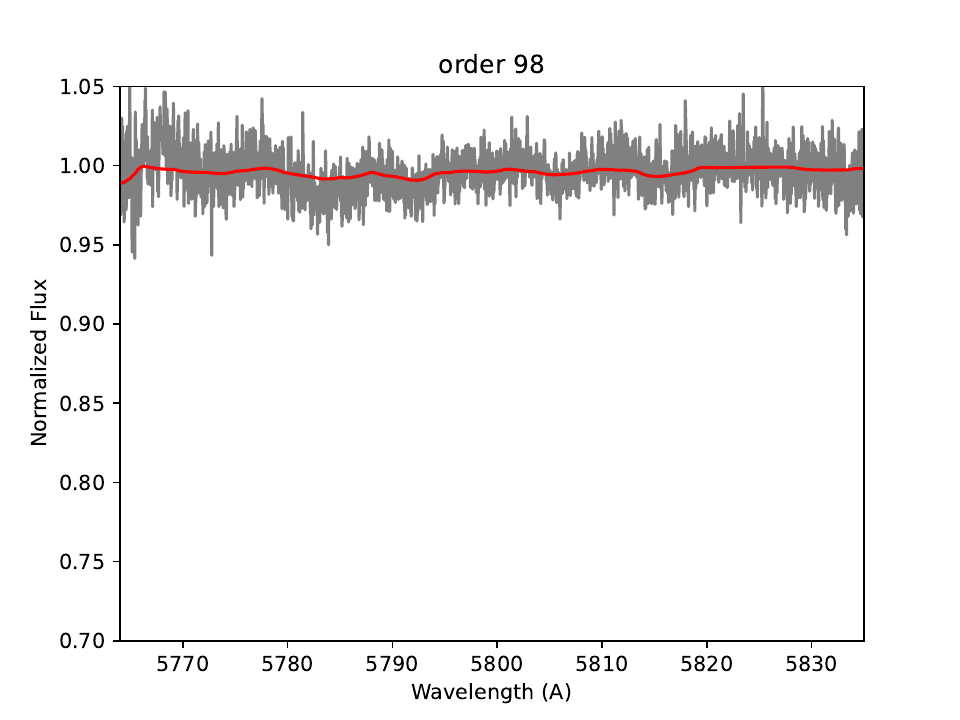}
        \caption{Order 98}
        \label{fig:s_ant_o98}
    \end{subfigure}
    \begin{subfigure}{0.32\textwidth}
        \noindent\includegraphics[width=\linewidth]{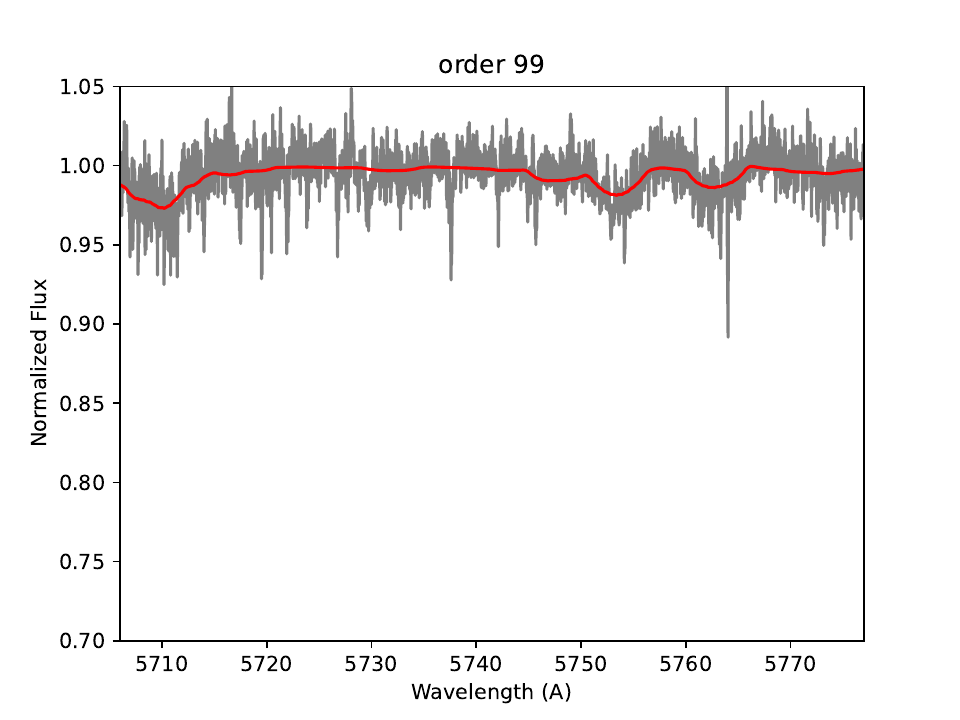}
        \caption{Order 99}
        \label{fig:s_ant_o99}
    \end{subfigure}
    \begin{subfigure}{0.32\textwidth}
        \includegraphics[width=\linewidth]{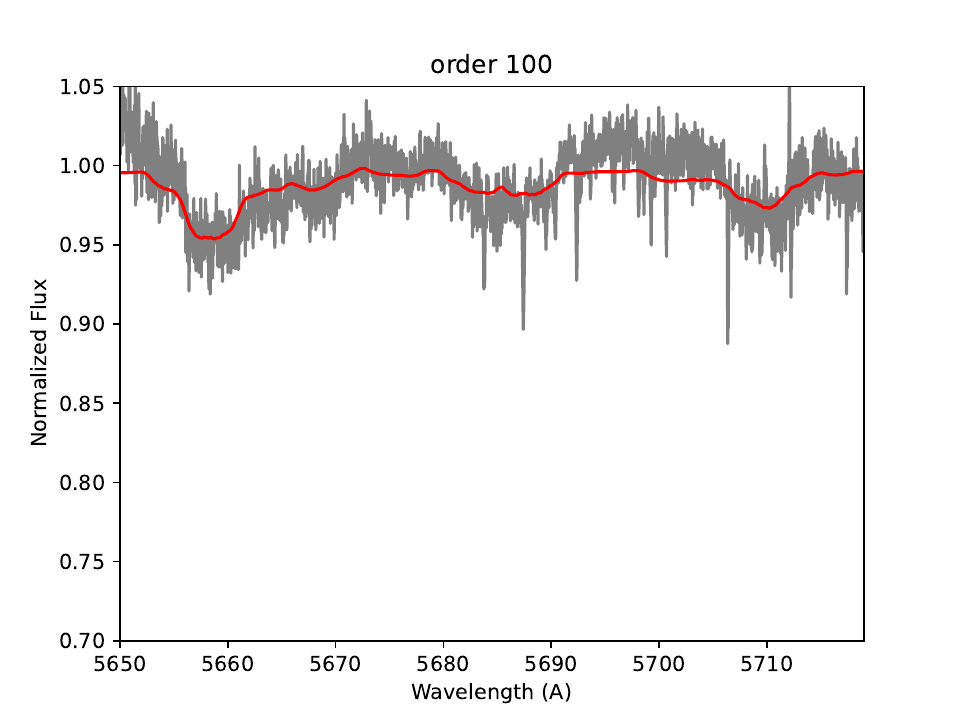}
        \caption{Order 100}
        \label{fig:s_ant_100}
    \end{subfigure}
    \begin{subfigure}{0.32\textwidth}
        \includegraphics[width=\linewidth]{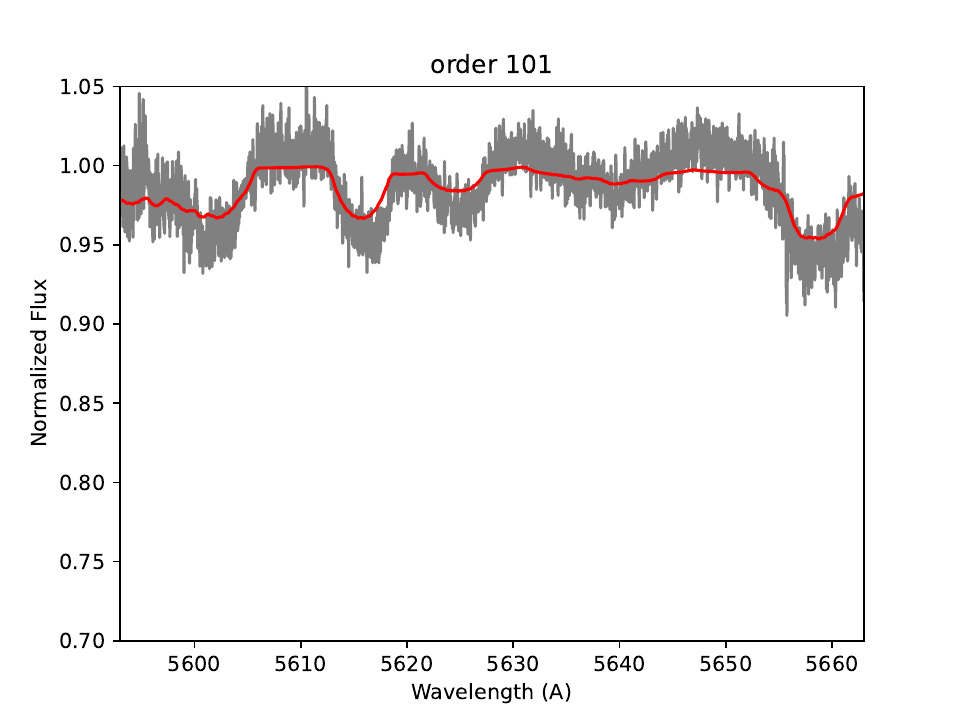}
        \caption{Order 101}
        \label{fig:s_ant_101}
    \end{subfigure}
    \begin{subfigure}{0.32\textwidth}
        \noindent\includegraphics[width=\linewidth]{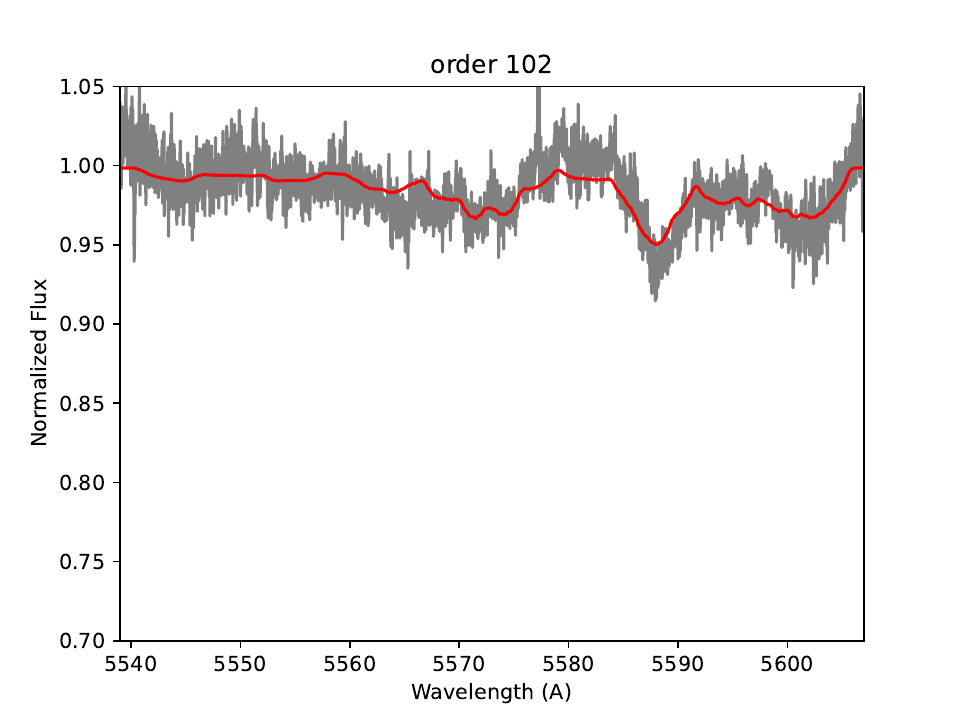}
        \caption{Order 102}
        \label{fig:s_ant_102}
    \end{subfigure}
    \begin{subfigure}{0.32\textwidth}
        \includegraphics[width=\linewidth]{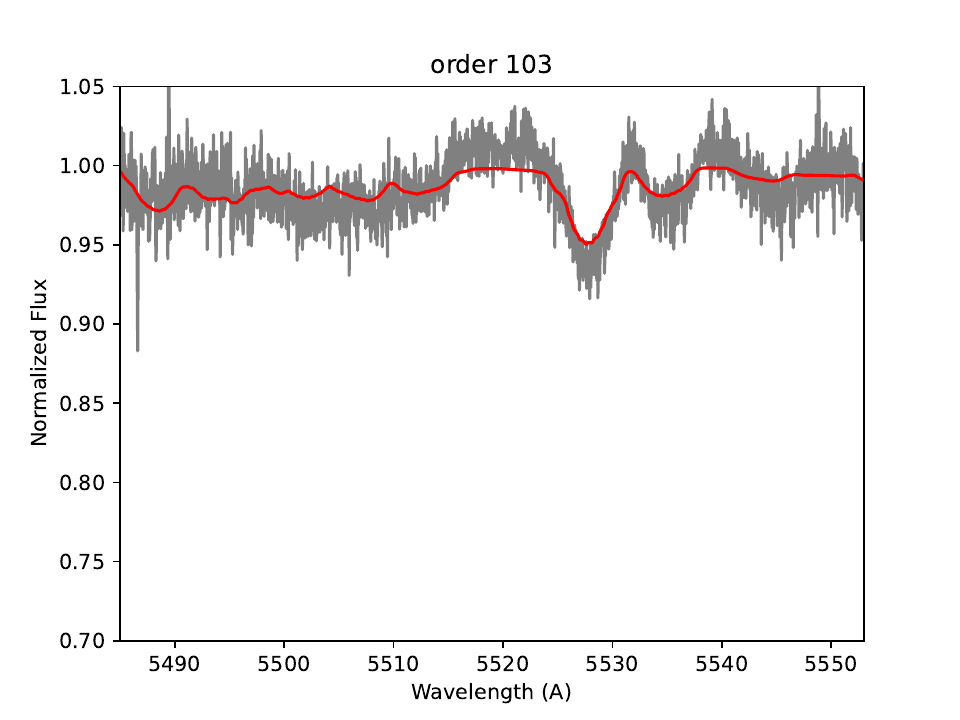}
        \caption{Order 103}
        \label{fig:s_ant_103}
    \end{subfigure}
    \begin{subfigure}{0.32\textwidth}
        \includegraphics[width=\linewidth]{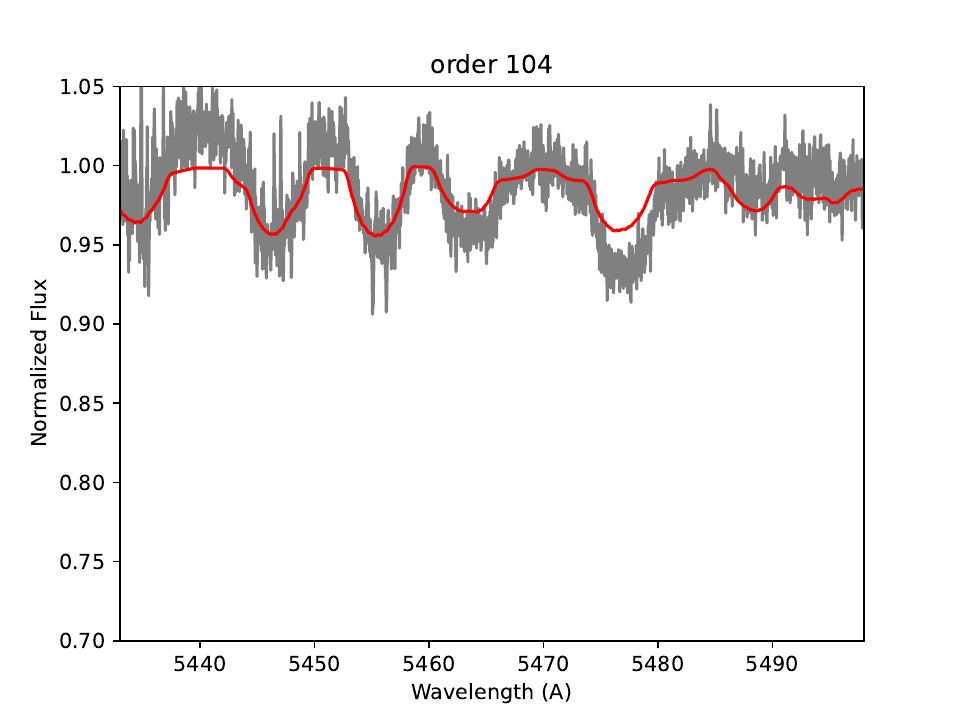}
        \caption{Order 104}
        \label{fig:s_ant_104}
    \end{subfigure}
    \begin{subfigure}{0.32\textwidth}
        \includegraphics[width=\linewidth]{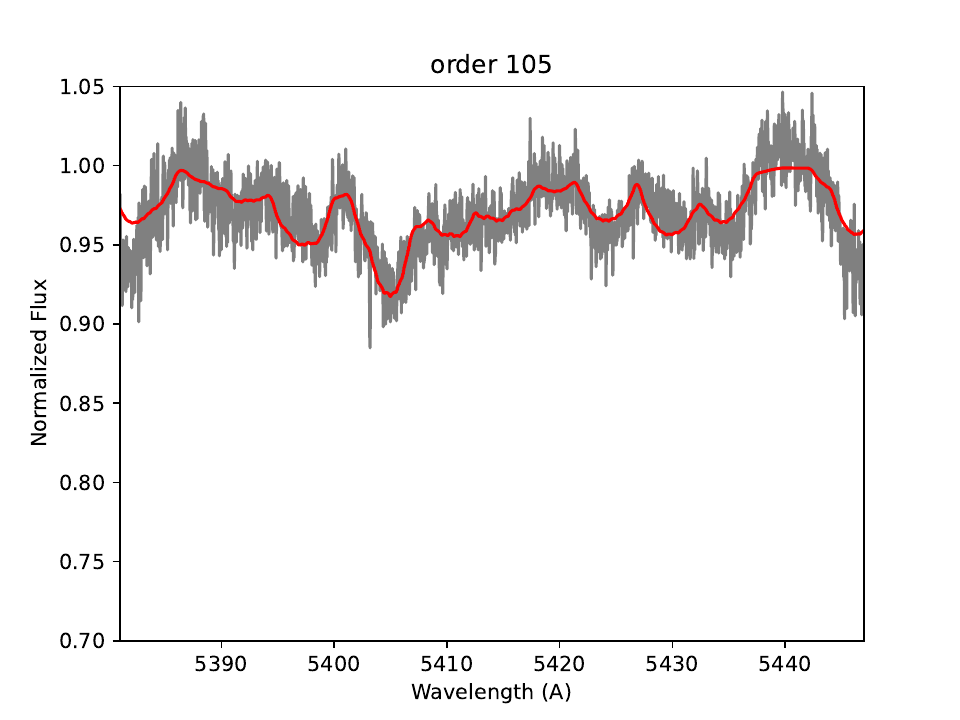}
        \caption{Order 105}
        \label{fig:s_ant_105}
    \end{subfigure}
    \begin{subfigure}{0.32\textwidth}
        \includegraphics[width=\linewidth]{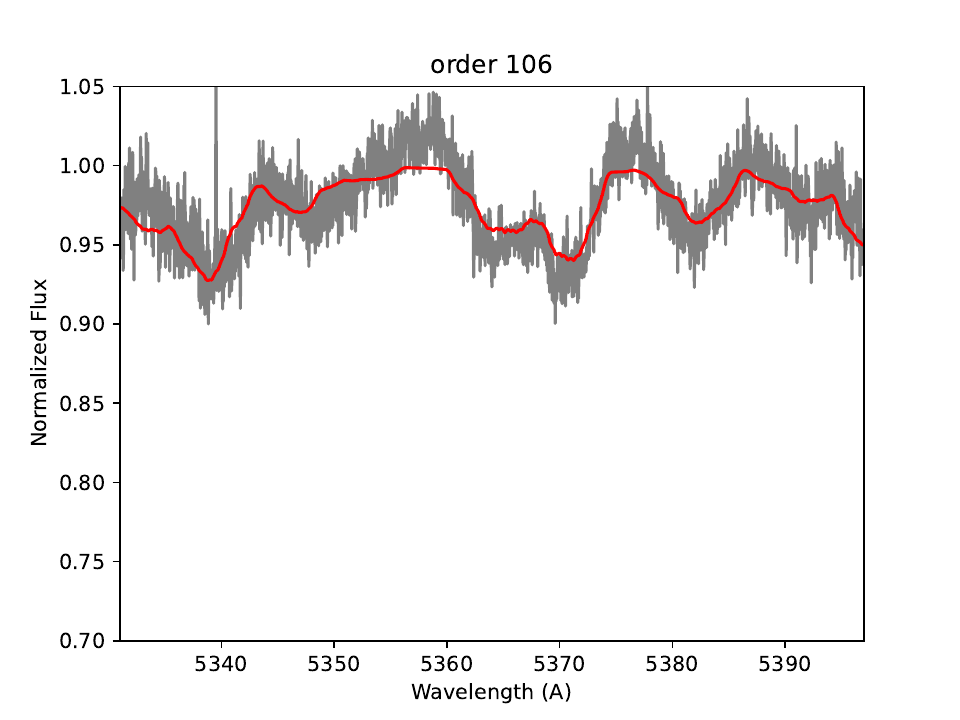}  
        \caption{Order 106}
        \label{fig:s_ant_106}
    \end{subfigure}
    \begin{subfigure}{0.32\textwidth}
        \includegraphics[width=\linewidth]{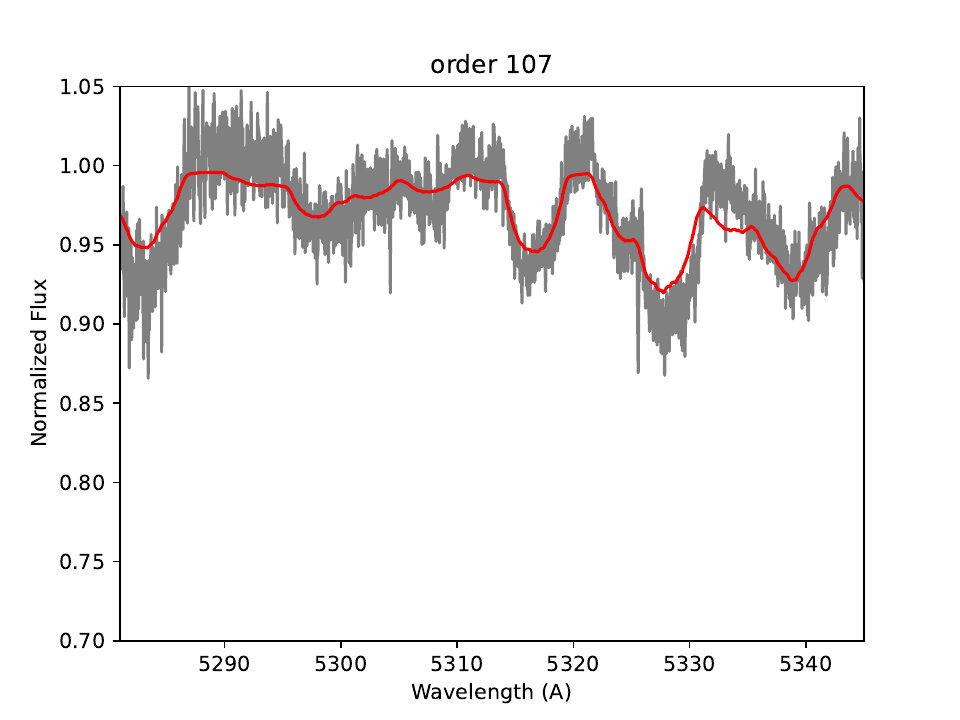} 
        \caption{Order 107}
        \label{fig:s_ant_107}
    \end{subfigure}    
     \caption{The observed spectrum (grey) of S Ant at the secondary (total) eclipse together with a Kurucz synthetic spectrum (red) calculated using solar abundances, with $T_{\rm eff}$ = 7100 K, log $g$=3.98 cgs, and $V_{\rm rot}$ sin$i$ = 150 km/s. The spectral order number is indicated in the sub-figure captions.
    \label{fig:SAnt_eclipse_spectrum}}
\end{figure*}

\begin{figure*}
    \centering %
    \begin{subfigure}{0.32\textwidth}
        \noindent\includegraphics[width=\linewidth]{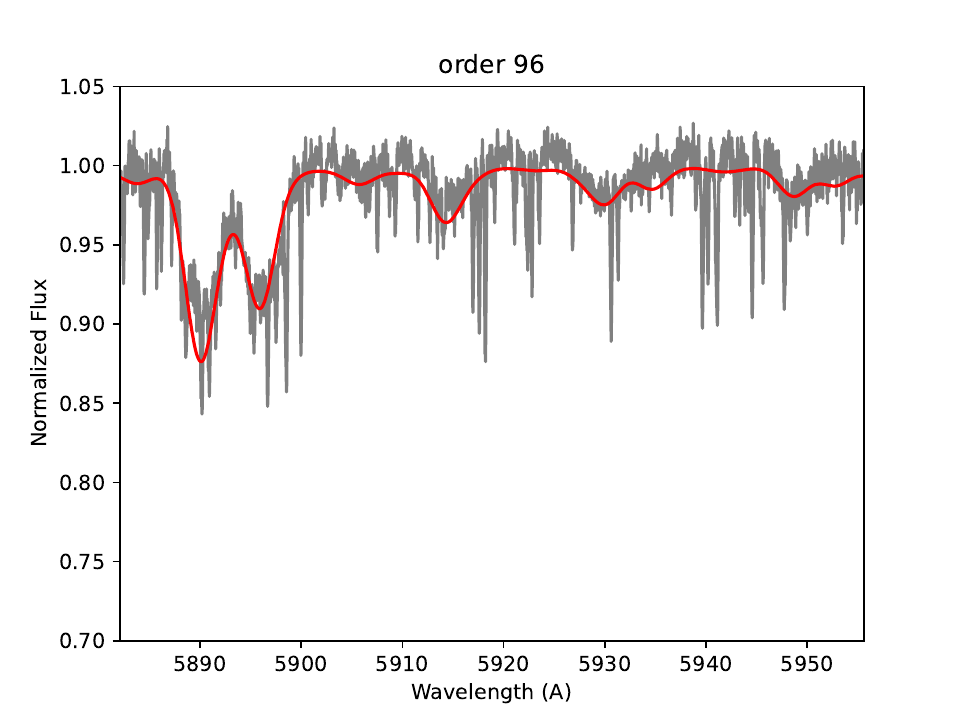}
        \caption{Order 96}
        \label{fig:eps_cra_o96}
    \end{subfigure}
    \begin{subfigure}{0.32\textwidth}
        \includegraphics[width=\linewidth]{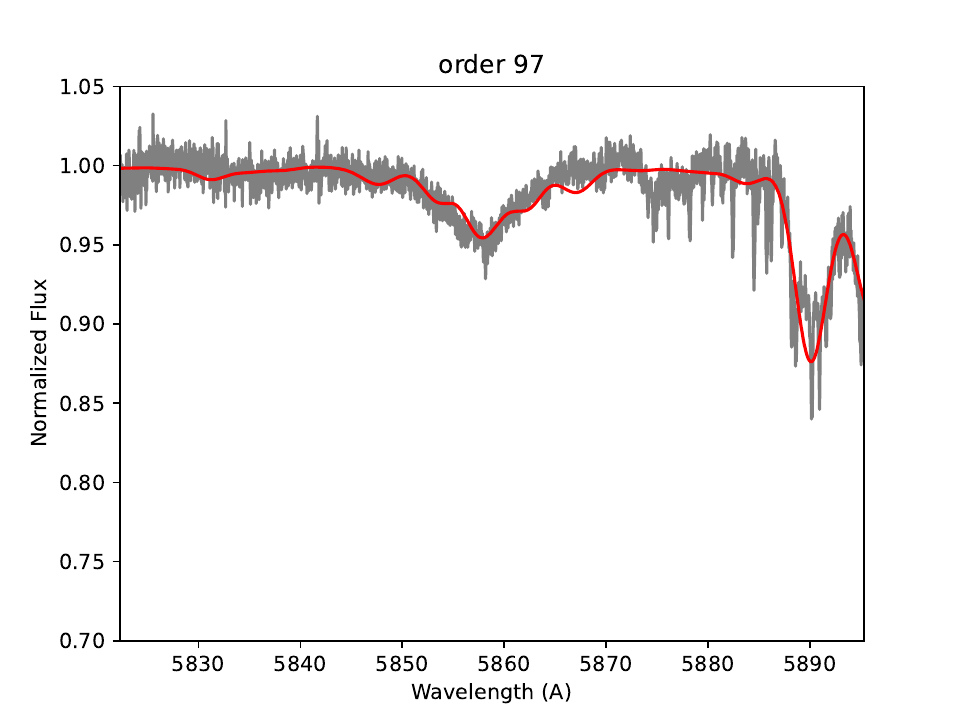}
        \caption{Order 97}
        \label{fig:eps_cra_o97}
    \end{subfigure}
    \begin{subfigure}{0.32\textwidth}
        \includegraphics[width=\linewidth]{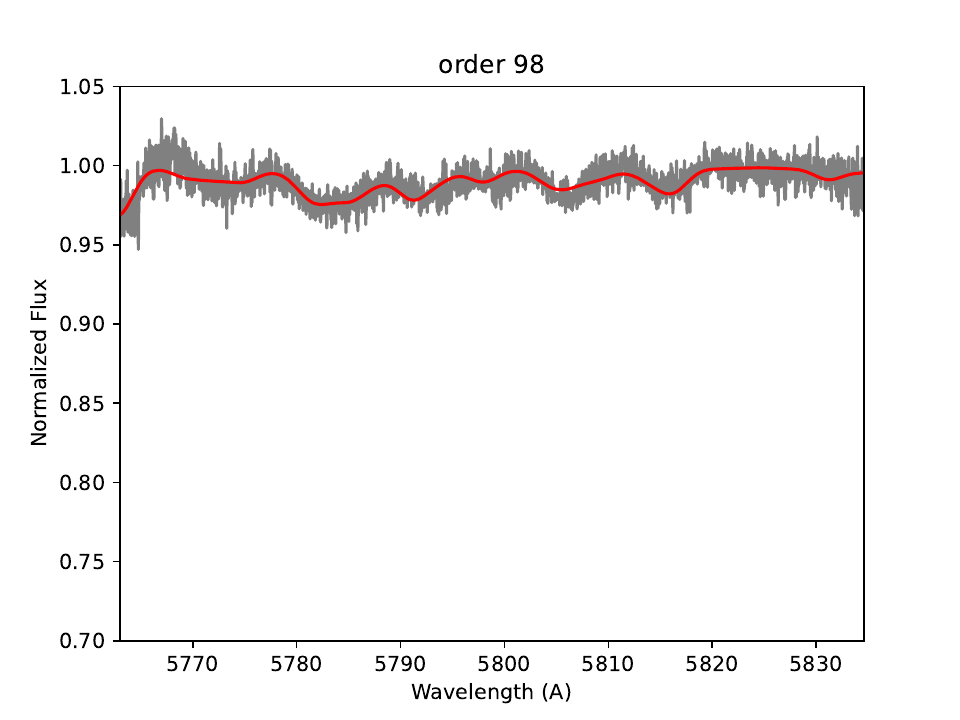}
        \caption{Order 98}
        \label{fig:eps_cra_o98}
    \end{subfigure}
    \begin{subfigure}{0.32\textwidth}
        \noindent\includegraphics[width=\linewidth]{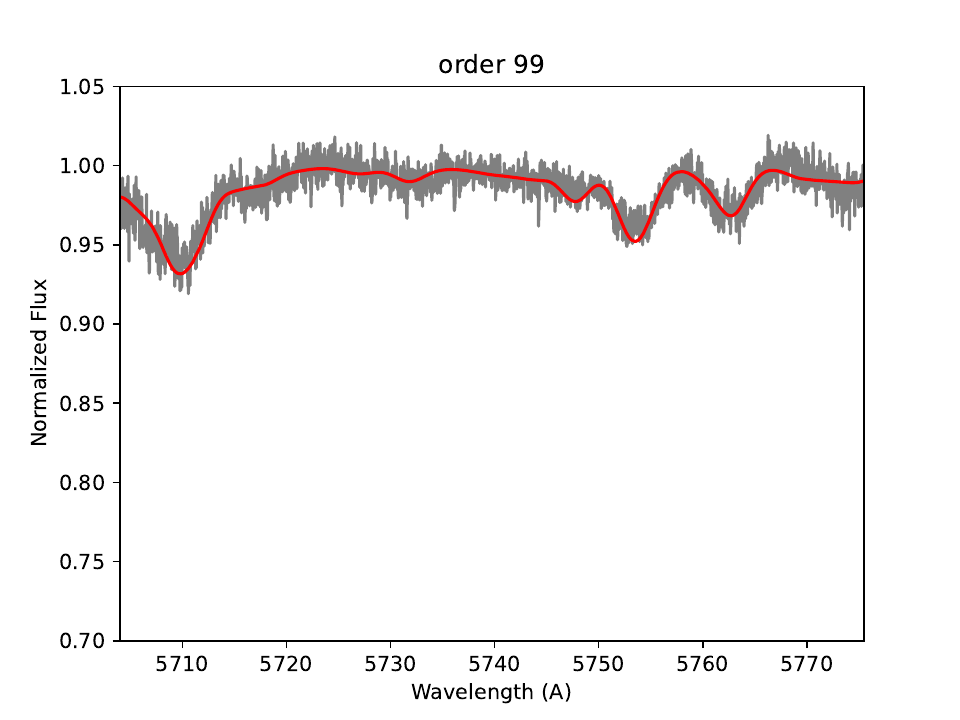}
        \caption{Order 99}
        \label{fig:eps_cra_o99}
    \end{subfigure}
    \begin{subfigure}{0.32\textwidth}
        \includegraphics[width=\linewidth]{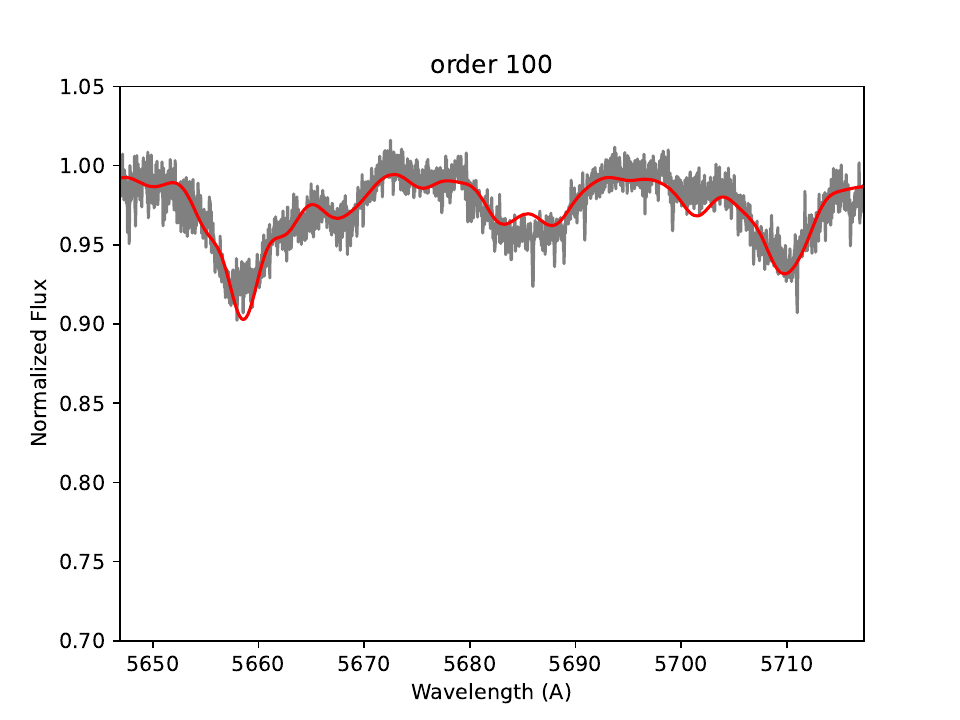}
        \caption{Order 100}
        \label{fig:eps_cra_100}
    \end{subfigure}
    \begin{subfigure}{0.32\textwidth}
        \includegraphics[width=\linewidth]{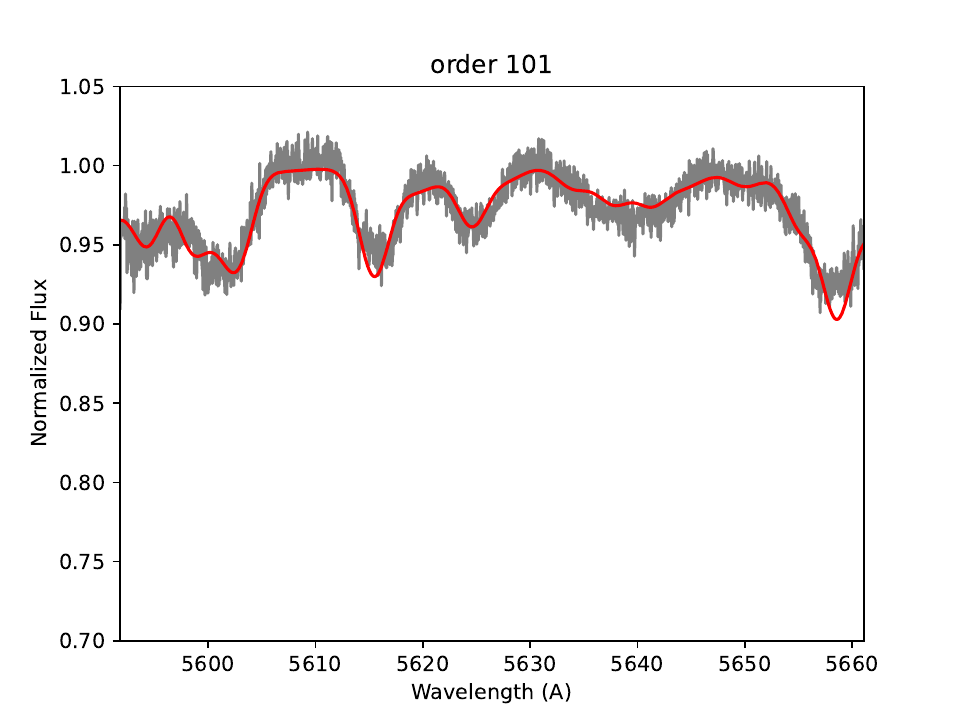}
        \caption{Order 101}
        \label{fig:eps_cra_101}
    \end{subfigure}
    \begin{subfigure}{0.32\textwidth}
        \noindent\includegraphics[width=\linewidth]{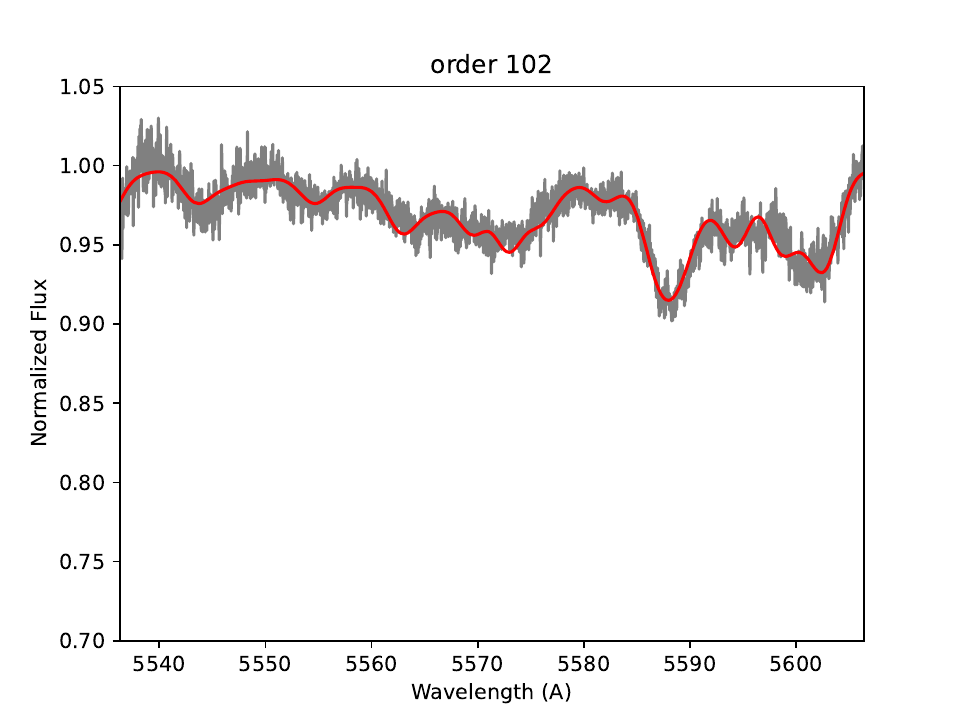}
        \caption{Order 102}
        \label{fig:eps_cra_102}
    \end{subfigure}
    \begin{subfigure}{0.32\textwidth}
        \includegraphics[width=\linewidth]{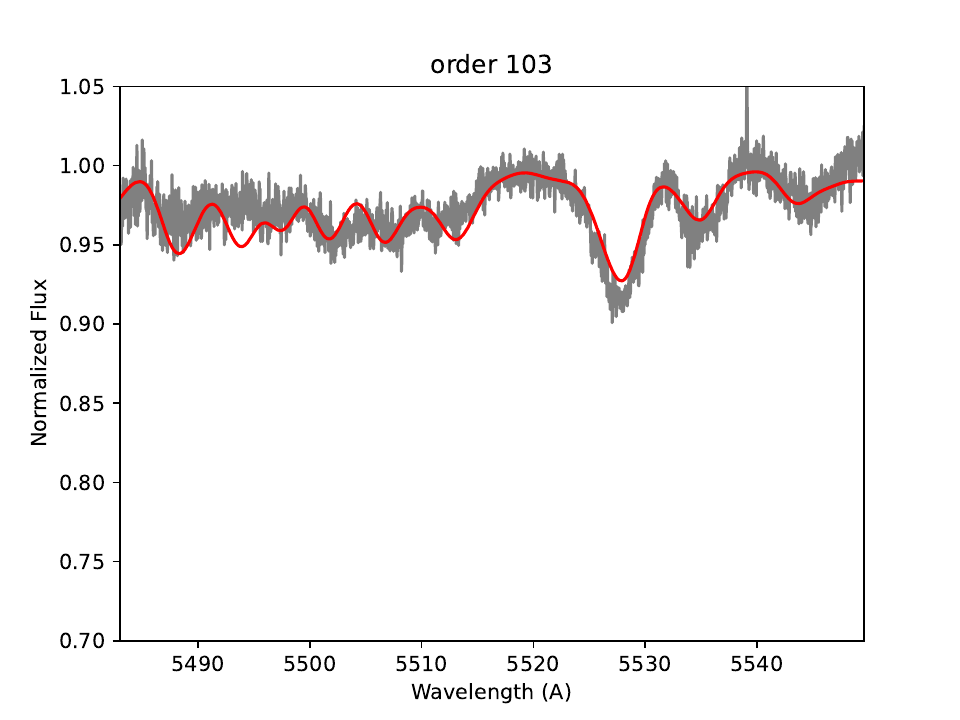}
        \caption{Order 103}
        \label{fig:eps_cra_103}
    \end{subfigure}
    \begin{subfigure}{0.32\textwidth}
        \includegraphics[width=\linewidth]{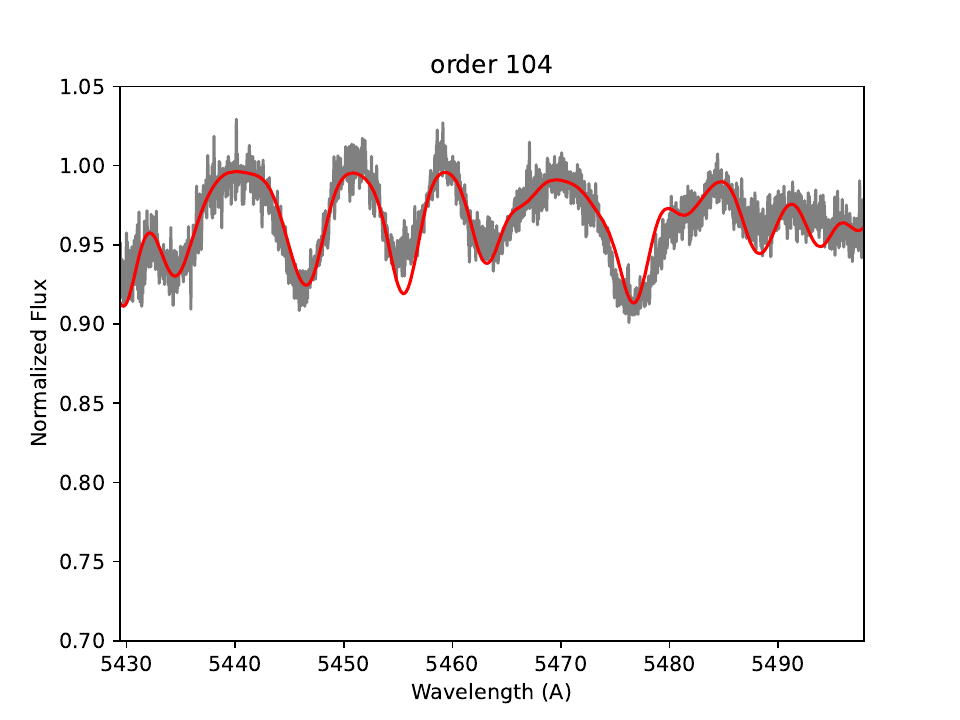}
        \caption{Order 104}
        \label{fig:eps_cra_104}
    \end{subfigure}
    \begin{subfigure}{0.32\textwidth}
        \includegraphics[width=\linewidth]{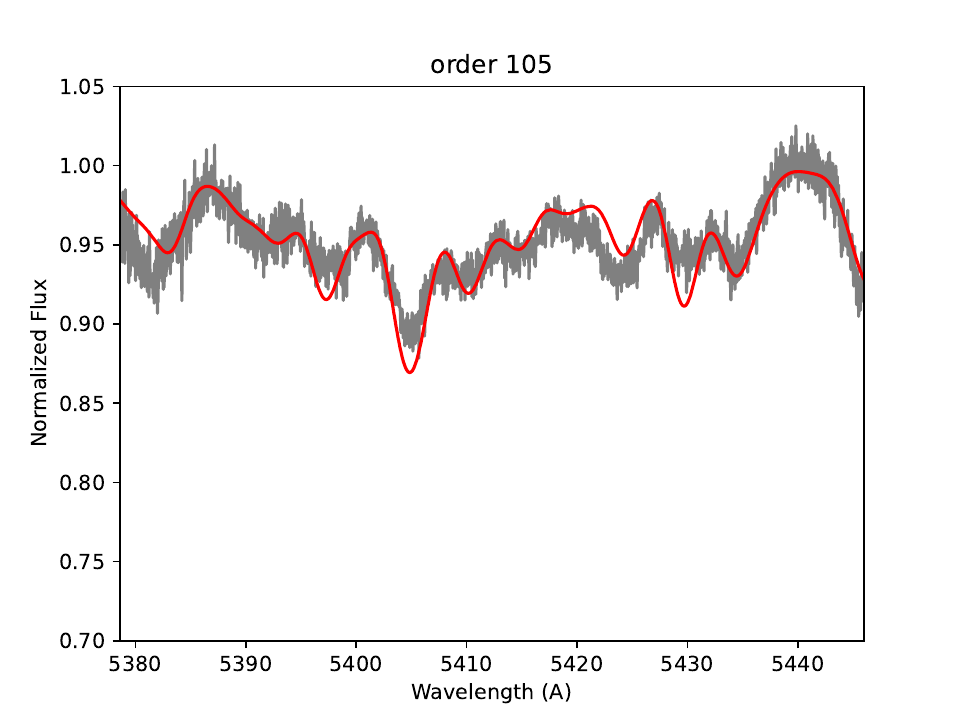}
        \caption{Order 105}
        \label{fig:eps_cra_105}
    \end{subfigure}
    \begin{subfigure}{0.32\textwidth}
        \includegraphics[width=\linewidth]{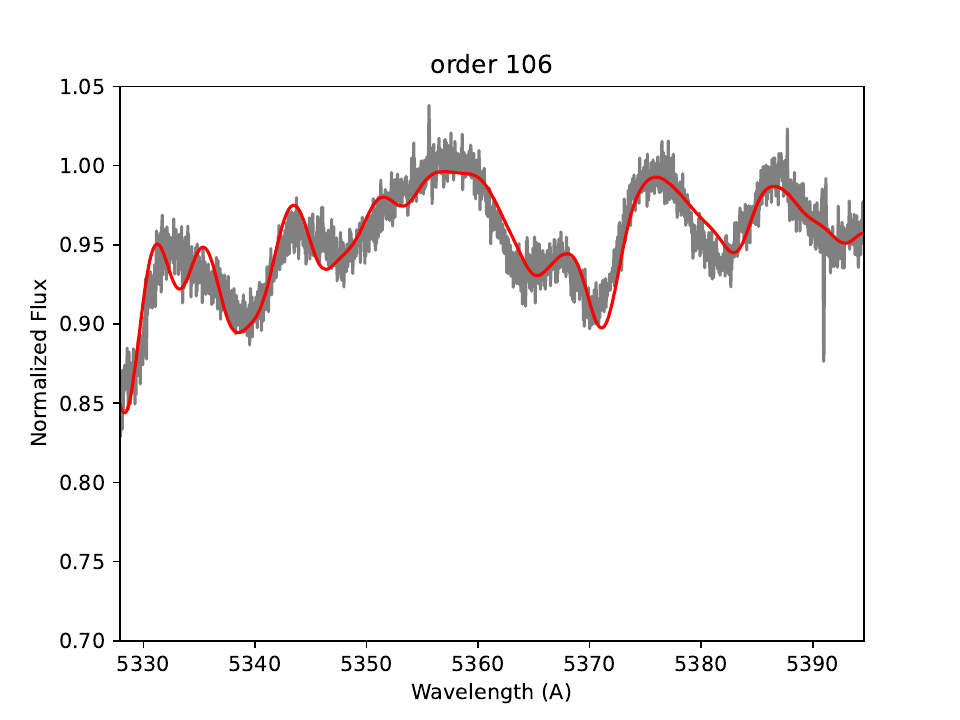}  
        \caption{Order 106}
        \label{fig:eps_cra_106}
    \end{subfigure}
    \begin{subfigure}{0.32\textwidth}
        \includegraphics[width=\linewidth]{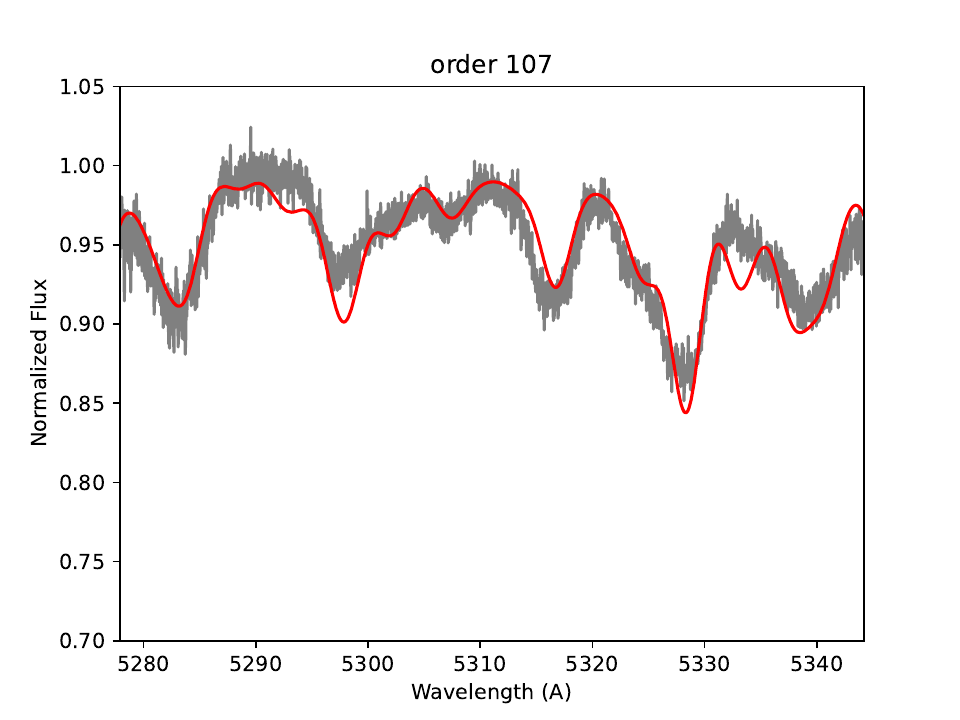} 
        \caption{Order 107}
        \label{fig:eps_cra_107}
    \end{subfigure}    
    \caption{The observed spectrum (grey) of $\epsilon$ CrA at the secondary (total) eclipse together with a Kurucz synthetic spectrum (red) calculated using solar abundances, with $T_{\rm eff}$ = 6000 K, log $g$=4.05 cgs, and $V_{\rm rot}$ sin$i$ =170 km/s. The spectral order number is indicated at the top of each panel.
    \label{fig:epsCrA_eclipse_spectrum}}
\end{figure*}

\label{lastpage}
\end{document}